\documentclass[pre,twocolumn,showpacs,groupaddress,preprintnumbers,floatfix]{revtex4-1}

\usepackage{etex}
\usepackage[mathscr]{eucal}
\usepackage{ifpdf}
\usepackage[pagebackref]{hyperref}
\usepackage{dcolumn}
\usepackage{url}
\usepackage{amsmath}
\usepackage{amscd}
\usepackage{amsfonts}
\usepackage{amssymb}
\usepackage{amsthm}
\usepackage{bigints}
\usepackage{xfrac}
\usepackage{braket}
\usepackage{bm}   
\usepackage{bbm}
\usepackage[abs]{overpic}
\usepackage{verbatim}
\usepackage{stmaryrd}
\usepackage{xcolor}
\usepackage{setspace}
\usepackage{tikz}
\usetikzlibrary{arrows}
\usetikzlibrary{automata}
\usetikzlibrary{positioning}
\usetikzlibrary{plotmarks}
\usepackage{pgfplots}
\pgfplotsset{compat=newest}

\usepackage{graphicx}
\usepackage{caption}
\usepackage{subcaption}

\usepackage{hyphenat}
\tikzstyle{vaucanson}=[
  node distance=2.5cm,
  bend angle=15,
  auto,
  every loop/.style={},
  every edge/.style={->,draw=black,line width=1.2,>=latex,shorten <=1pt,shorten >=1pt},
  every state/.style={draw=black,line width=2,font=\large},
  loop right/.style={right,out=22,in=-22,loop},
  loop above/.style={above,out=112,in=68,loop},
  loop left/.style={left,out=202,in=158,loop},
  loop below/.style={below,out=292,in=248,loop},
  loop above right/.style={right,out=67,in=23,loop},
  loop above left/.style={left,out=157,in=113,loop},
  loop below left/.style={left,out=247,in=203,loop},
  loop below right/.style={right,out=337,in=293,loop},
]

\theoremstyle{plain}    
\theoremstyle{plain}    
\theoremstyle{plain}    
\theoremstyle{plain}    
\theoremstyle{plain}    \newtheorem{The}{Theorem}
\theoremstyle{plain}    \newtheorem*{ProThe}{Proof}
\theoremstyle{plain}    
\theoremstyle{plain}    
\theoremstyle{plain}    
\theoremstyle{plain}    
\theoremstyle{plain}    \newtheorem{Def}{Definition}
\theoremstyle{plain}    
\theoremstyle{plain}

\newcommand{\MeasAlphabet}  {\mathcal{A}}
\newcommand{\MeasSymbol}   { {X} }
\newcommand{\meassymbol}   { {x} }

\newcommand{\CausalState}   { \mathcal{S} }

\newcommand{\causalstate}   { \sigma }
\newcommand{\CausalStateSet}    { \boldsymbol{\CausalState} }
\newcommand{\AlternateState}    { \mathcal{R} }

\newcommand{\AlternateStateSet} { \boldsymbol{\AlternateState} }

\newcommand{\EE}        {{\bf E}}

\newcommand{\PC}        {\chi}

\newcommand{\ProcessAlphabet}   {\MeasAlphabet}

\newcommand{\forward}{+}
\newcommand{\reverse}{-}
\newcommand{\forwardreverse}{\pm} 

\newcommand{\FutureCausalState} { {\CausalState}^{\forward} }

\newcommand{\PastCausalState}   { {\CausalState}^{\reverse} }

\newcommand{\one}{\mathbf{1}}

\newcommand{\lastindex}[2]{
  \edef\tempa{0}
  \edef\tempb{#2}
  \ifx\tempa\tempb
    \edef\tempc{#1}
  \else
    \edef\tempa{0}
    \edef\tempb{#1}
    \ifx\tempa\tempb
      \edef\tempc{#2}
    \else
      \edef\tempc{#1+#2}
    \fi
  \fi
  \tempc
}

\newcommand{\CSjoint}[1][,]{
   \edef\tempa{:}
   \edef\tempb{#1}
   \ifx\tempa\tempb
      \ensuremath{\FutureCausalState\!#1\PastCausalState}
   \else
      \ensuremath{\FutureCausalState#1\PastCausalState}
   \fi
}

\newif\ifpm
\edef\tempa{\forwardreverse}
\edef\tempb{\pm}
\ifx\tempa\tempb
   \pmfalse
\else
   \pmtrue
\fi

\newcommand{\z}{z}  
\newcommand{\R}{R}  

\colorlet {R_color}    {blue}
\colorlet {k_color}    {black!30!green}

\def\clap#1{\hbox to 0pt{\hss#1\hss}}

\begin{document}

\title{Beyond the Spectral Theorem:\\
\vspace{0.05in}
Spectrally Decomposing Arbitrary Functions of\\
Nondiagonalizable Operators}

\author{Paul M. Riechers}
\email{pmriechers@ucdavis.edu}

\author{James P. Crutchfield}
\email{chaos@ucdavis.edu}

\affiliation{Complexity Sciences Center\\
Department of Physics\\
University of California at Davis\\
One Shields Avenue, Davis, CA 95616}

\date{\today}
\bibliographystyle{unsrt}

\begin{abstract}
Nonlinearities in finite dimensions can be linearized by projecting them into
infinite dimensions. Unfortunately, often the linear operator techniques
that one would then use simply fail since the operators cannot be diagonalized.
This curse is well known. It also occurs for finite-dimensional linear
operators. We circumvent it by developing a \emph{meromorphic functional
calculus} that can decompose arbitrary functions of nondiagonalizable linear
operators in terms of their eigenvalues and projection operators. It extends
the spectral theorem of normal operators to a much wider class, including
circumstances in which poles and zeros of the function coincide with the
operator spectrum. By allowing the direct manipulation of individual
eigenspaces of nonnormal and nondiagonalizable operators, the new theory avoids
spurious divergences. As such, it yields novel insights and closed-form
expressions across several areas of physics in which nondiagonalizable dynamics
are relevant, including memoryful stochastic processes, open nonunitary quantum
systems, and far-from-equilibrium thermodynamics.

The technical contributions include the first full treatment of arbitrary
powers of an operator. In particular, we show that the Drazin inverse,
previously only defined axiomatically, can be derived as the negative-one power
of singular operators within the meromorphic functional calculus and we give a
general method to construct it. We provide new formulae for constructing
projection operators and delineate the relations between projection operators,
eigenvectors, and generalized eigenvectors.

By way of illustrating its application, we explore several, rather distinct
examples. First, we analyze stochastic transition operators in discrete and
continuous time. Second, we show that nondiagonalizability can be a robust
feature of a stochastic process, induced even by simple counting. As
a result, we directly derive distributions of the Poisson process and point out
that nondiagonalizability is intrinsic to it and the broad class of hidden
semi-Markov processes. Third, we show that the Drazin inverse arises naturally
in stochastic thermodynamics and that applying the meromorphic functional
calculus provides closed-form solutions for the dynamics of key thermodynamic
observables. Fourth, we show that many memoryful processes have power spectra
indistinguishable from white noise, despite being highly organized.
Nevertheless, whenever the power spectrum is nontrivial, it is a direct
signature of the spectrum and projection operators of the process' hidden
linear dynamic, with nondiagonalizable subspaces yielding qualitatively
distinct line profiles. Finally, we draw connections to the
Ruelle--Frobenius--Perron and Koopman operators for chaotic dynamical systems.
\end{abstract}



\preprint{Santa Fe Institute Working Paper 16-07-015}
\preprint{arxiv.org:1607.06526 [math-ph]}

\maketitle



\setstretch{1.1}

\newcommand{\pdf}{\text{p}}
\newcommand{\Abet}{\ProcessAlphabet}
\newcommand{\MS}{\MeasSymbol}
\newcommand{\ms}{\meassymbol}
\newcommand{\SSet}{\CausalStateSet}
\newcommand{\St}{\CausalState}
\newcommand{\st}{s}
\newcommand{\cs}{\causalstate}
\newcommand{\syncMSP}{{$\mathscr{S}$-MSP}}
\newcommand{\crypticMSP}{{$\PC$-MSP}}
\newcommand{\MxSt}{\AlternateState}
\newcommand{\MxSSet}{\AlternateStateSet_\pi}
\newcommand{\mxst}{\eta}
\newcommand{\mxstw}[1]{\mxst_{#1}} 		
\newcommand{\StartMS}{\bra{\delta_\pi}}
\newcommand{\opGen}{A}         
\newcommand{\matHMM}{T}     
\newcommand{\matMSP}{W}     
\newcommand{\TentT}{\varTheta}

\newcommand{\HWA}{\ket{H(W^\Abet)}}
\newcommand{\Hmxst}{\ket{H[\mxst]}}
\newcommand{\hsym}{\reflectbox{h}\text{h}}
\newcommand{\Redundancy}{\boldsymbol{R}}
\newcommand{\ACEphemeral}{\gamma_{\multimap}}
\newcommand{\hEphemeral}{h_{\multimap}}
\newcommand{\ACPersistent}{\gamma_{\rightsquigarrow}}
\newcommand{\hPersistent}{h_{\rightsquigarrow}}
\newcommand{\EEEphemeral}{\EE_{\multimap}}
\newcommand{\EEPersistent}{\EE_{\rightsquigarrow}}

\newcommand{\corrbra}{\bra{\pi \overline{\Abet}}}
\newcommand{\corrket}{\ket{\Abet \one}}

\newcommand{\Cmatrix}{\mathcal{C}}   
\newcommand{\LWwoutZero}{\Lambda_W^{\setminus 0}}

\newcommand{\distr}{\boldsymbol{\mu}}
\newcommand{\rate}{G}
\newcommand{\stationary}{\boldsymbol{\pi}}
\newcommand{\EnvSSet}{\boldsymbol{\mathcal{V}}}
\newcommand{\control}{\vec{v}}
\newcommand{\Gcontrol}{G^{(\SSet \to \SSet | \control)}}
\newcommand{\JointSSet}{\boldsymbol{\mathcal{U}}}

\newcommand{\Heff}{H^\text{eff}}
\newcommand{\Qex}{Q_\text{ex}}
\newcommand{\Qhk}{Q_\text{hk}}
\newcommand{\Wex}{W_\text{ex}}
\newcommand{\kB}{k_\text{B}}
\newcommand{\eeep}{\omega_\text{ex}}
\newcommand{\Sss}{S^\text{ss}_{\control}}
\newcommand{\dSq}{\dot{\varPi}}

\begin{quote}
\emph{... the supreme goal of all
theory is to make the irreducible basic elements as simple and as few as
possible without having to surrender the adequate representation of a single
datum of experience.}
~~~~~~~~~~~A. Einstein \cite[p. 165]{Eins34a}
\end{quote}

\section{Introduction}

Decomposing a complicated system into its constituent parts---reductionism---is
one of science's most powerful strategies for analysis and understanding.
Large-scale systems with linearly coupled components give one paradigm of this
success. Each can be decomposed into an equivalent system of independent
elements using a similarity transformation calculated by the linear algebra of
the system's eigenvalues and eigenvectors. The physics of linear wave
phenomena, whether of classical light or quantum mechanical amplitudes, sets
the standard of complete reduction rather high. The dynamics is captured by an
``operator'' whose allowed or exhibited ``modes'' are the elementary behaviors
out of which composite behaviors are constructed by simply weighing each mode's
contribution and adding them up. 

However, one should not reduce a composite system more than is necessary nor,
as is increasingly appreciated these days, more than one, in fact, can. Indeed,
we live in a complex, nonlinear world whose constituents are strongly
interacting. Often their key structures and memoryful behaviors emerge only
over space and time. These are the complex systems. Yet, perhaps surprisingly,
many complex systems with nonlinear dynamics correspond to linear operators in
abstract high-dimensional spaces~\cite{Koop31, Gasp95, Budi12}. And so, there
is a sense in which even these complex systems can be reduced to the study of
independent nonlocal collective modes.

Reductionism, however, faces its own challenges even within its paradigmatic
setting of linear systems: linear operators may have interdependent modes with
irreducibly entwined behaviors. These irreducible components correspond to
so-called nondiagonalizable subspaces. No similarity transformation can reduce
them.

In this view, reductionism can only ever be a guide. The actual goal is to
achieve a happy medium, as Einstein reminds us, of decomposing a system only to
that level at which the parts are irreducible. 
To proceed, though, begs the
original question, What happens when reductionism fails? To answer this
requires revisiting one of its more successful implementations, spectral
decomposition of completely reducible operators.

\subsection{Spectral Decomposition}
\label{sec:SpecDecompReview}

Spectral decomposition---splitting a linear operator into independent modes of
simple behavior---has greatly accelerated progress in the physical sciences.
The impact stems from the fact that spectral decomposition is not only a
powerful mathematical tool for expressing the organization of large-scale
systems, but also yields predictive theories with directly observable physical
consequences~\cite{Tref11}. Quantum mechanics and statistical mechanics
identify the energy eigenvalues of Hamiltonians as the basic objects in
thermodynamics: transitions among the energy eigenstates yield heat and work.
The spectrum of eigenvalues reveals itself most directly in other kinds of
spectra, such as the frequency spectra of light emitted by the gases that
permeate the galactic filaments of our universe~\cite{Sand76}. Quantized
transitions, an initially mystifying feature of atomic-scale systems,
correspond to distinct eigenvectors and discrete spacing between eigenvalues.
The corresponding theory of spectral decomposition established the quantitative
foundation of quantum mechanics.

The applications and discoveries enabled by spectral decomposition and the
corresponding spectral theory fill a long list. In application, direct-bandgap
semiconducting materials can be turned into light-emitting diodes (LEDs) or
lasers by engineering the spatially-inhomogeneous distribution of energy
eigenvalues and the occupation of their corresponding states~\cite{Miln86}.
Before their experimental discovery, anti-particles were anticipated
as the nonoccupancy of negative-energy eigenstates of the Dirac
Hamiltonian~\cite{Dira33a}.

The spectral theory, though, extends far beyond physical science disciplines.
In large measure, this arises since the evolution of any object corresponds to
a linear dynamic in a sufficiently high-dimensional state space. Even nominally
nonlinear dynamics over several variables, the canonical mechanism of
deterministic chaos, appear as linear dynamics in appropriate
infinite-dimensional shift-spaces~\cite{Budi12}. A nondynamic version of
rendering nonlinearities into linearities in a higher-dimensional feature space
is exploited with much success today in machine learning by support vector
machines, for example~\cite{Cort95}. Spectral decomposition often allows a
problem to be simplified by approximations that use only the dominant
contributing modes. Indeed, human-face recognition can be efficiently
accomplished using a small basis of ``eigenfaces''~\cite{Siro87}.

Certainly, there are many applications that highlight the importance of
decomposition and the spectral theory of operators. However, a brief reflection
on the mathematical history will give better context to its precise results,
associated assumptions, and, more to the point, the generalizations we develop
here in hopes of advancing the analysis and understanding of complex systems.

Following on early developments of operator theory  by Hilbert and
co-workers~\cite{Cour53}, the \emph{spectral theorem for normal operators}
reached maturity under von Neumann by the early 1930s~\cite{Neum30,Neum55}. It
became the mathematical backbone of much progress in physics since then, from
classical partial differential equations to quantum physics. Normal operators,
by definition, commute with their Hermitian conjugate: $A^\dagger A = A
A^\dagger$. Examples include symmetric and orthogonal matrices in classical
mechanics and Hermitian, skew-Hermitian, and unitary operators in quantum
mechanics.

The spectral theorem itself is often identified as a collection of
related results about normal operators; see, e.g., Ref.~\cite{Hass99a}. In the
case of finite-dimensional vector spaces \cite{Halm58a}, the spectral theorem
asserts that normal operators are diagonalizable and can always be diagonalized
by a unitary transformation; that left and right eigenvectors (or
eigenfunctions) are simply related by complex-conjugate transpose; that these
eigenvectors form a complete basis; and that functions of a normal operator
reduce to the action of the function on each eigenvalue. Most of these
qualities survive with only moderate provisos in the infinite-dimensional case.
In short, the spectral theorem makes physics governed by normal operators
tractable.

The spectral theorem, though, appears powerless when faced with nonnormal and
nondiagonalizable operators. What then are we to do when confronted, say, by
complex interconnected systems with nonunitary time evolution, by open systems,
by structures that emerge on space and time scales different from the equations
of motion, or by other frontiers of physics governed by nonnormal and
not-necessarily-diagonalizable operators? Where is the comparably constructive
framework for calculations beyond the standard spectral theorem? Fortunately,
portions of the necessary generalization have been made within pure
mathematics~\cite{Dunf54a}, some finding applications in engineering and
control~\cite{Meyer00,Ants07}. However, what is available is incomplete. And,
even that which is available is often not in a form adapted to perform
calculations that lead to quantitative predictions.

\subsection{Synopsis}
\label{sec:Synopsis}

Here, we build on previous work in functional analysis and operator theory to
provide both a rigorous and constructive foundation for physically relevant
calculations involving not-necessarily-diagonalizable operators. In effect, we
extend the spectral theorem for normal operators to a broader setting, allowing
generalized ``modes'' of nondiagonalizable systems to be identified and
manipulated. The meromorphic functional calculus we develop extends Taylor
series expansion and standard holomorphic functional calculus to analyze
arbitrary functions of not-necessarily-diagonalizable operators. It readily
handles singularities arising when poles (or zeros) of the function coincide
with poles of the operator's resolvent---poles that appear precisely at the
eigenvalues of the operator. Pole--pole and pole--zero interactions
substantially modify the complex-analytic residues within the functional
calculus. A key result is that the negative-one power of a singular operator
exists in the meromorphic functional calculus. It is the \emph{Drazin inverse},
a powerful tool that is receiving increased attention in stochastic
thermodynamics.

Taken altogether, the functional calculus, Drazin inverse, and methods to
manipulate particular eigenspaces, are key to a thorough-going analysis of many
complex systems, many now accessible for the first time. Indeed, the framework
has already been fruitfully employed by the authors in several specific
applications, including closed-form expressions for signal processing and
information measures of hidden Markov processes \cite{Crut13a,Riec14b,Riec14a}
and for compressing stochastic processes over a quantum channel
\cite{Riec16a}.  However, the techniques are sufficiently general they will be
much more widely useful. We envision new opportunities for similar detailed
analyses, ranging from biophysics to quantum field theory, wherever
restrictions to normal operators and diagonalizability have been roadblocks.

With this broad scope in mind, we develop the mathematical theory first without
reference to specific applications and disciplinary terminology. We later give
pedagogical (yet, we hope, interesting) examples, exploring several niche, but
important applications to finite hidden Markov processes, basic stochastic
process theory, nonequilibrium thermodynamics, signal processing, and nonlinear
dynamical systems. At a minimum, the examples and their breadth serve to better
acquaint readers with the basic methods required to employ the theory.

We introduce the meromorphic functional calculus in \S
\ref{sec:FunctionalCalculi} through \S \ref{sec:EverythingDecomposed}, after
necessary preparation in \S \ref{sec:SpectralPrimer}. \S
\ref{sec:IndexOneProjOps} further explores eigenprojectors, which we refer to
here simply as \emph{projection operators}. \S
\ref{sec:ProjectorsAndEigenvectors} makes explicit their relationship
with eigenvectors and generalized eigenvectors. \S
\ref{sec:MFC_in_special_cases} then discusses simplifications of the functional
calculus for special cases, while \S \ref{sec:StochasticSpectra} takes up the
spectral properties of transition operators. The examples are discussed at
length in \S \ref{sec:Examples} before we close in \S \ref{sec:Conclusion}
with suggestions on future applications and research directions.

\section{Spectral Primer}
\label{sec:SpectralPrimer}

The following is relatively self-contained, assuming basic familiarity with
linear algebra at the level of Refs.~\cite{Halm58a,Meyer00}---including
eigen-decomposition and knowledge of the Jordan canonical form, partial
fraction expansion (see Ref.~\cite{Latni98}), and series expansion---and basic
knowledge of complex analysis---including the residue theorem and calculation
of residues at the level of Ref.~\cite{Boas66}.  For those lacking a working
facility with these concepts, a quick review of \S \ref{sec:Examples}'s
applications may motivate reviewing them.  In this section, we introduce our
notation and, in doing so, remind the reader of certain basic concepts in
linear algebra and complex analysis that will be used extensively in the following.

To begin, we restrict attention to operators with finite representations and
only sometimes do we take the limit of dimension going to infinity. That is, we
do not consider infinite-rank operators outright. While this runs counter to
previous presentations in mathematical physics that consider only
infinite-dimensional operators, the upshot is that they---as limiting
operators---can be fully treated with a countable point spectrum. We present
examples of this later on. Accordingly, we restrict our attention to operators
with at most a countably infinite spectrum. Such operators share many features
with finite-dimensional square matrices, and so we recall several elementary
but essential facts from matrix theory used extensively in the main development.

If $\opGen$ is a finite-dimensional square matrix, then its \emph{spectrum} is
simply the set $\Lambda_\opGen$ of its eigenvalues:
\begin{align*}
\Lambda_\opGen = \bigl\{ \lambda \in \mathbb{C}: \text{det}(\lambda I - \opGen) = 0 \bigr\}
  ~,
\end{align*}
where det$(\cdot)$ is the determinant of its argument and $I$ is the identity
matrix. The \emph{algebraic multiplicity} $a_\lambda$ of eigenvalue $\lambda$
is the power of the term $(z-\lambda)$ in the characteristic polynomial
det$(zI - \opGen)$. In contrast, the \emph{geometric multiplicity} $g_\lambda$
is the dimension of the kernel of the transformation $\opGen - \lambda I$ or,
equivalently, the number of linearly independent eigenvectors associated with
the eigenvalue. The algebraic and geometric multiplicities are all equal when
the matrix is diagonalizable.

Since there can be multiple subspaces associated with a single eigenvalue,
corresponding to different Jordan blocks in the Jordan canonical form, it is
structurally important to distinguish the \emph{index} of the eigenvalue
associated with the largest of these subspaces~\cite{Dunf43a}.

\begin{Def} 
Eigenvalue $\lambda$'s \emph{index} $\nu_\lambda$ is the size of the largest
Jordan block associated with $\lambda$.  
\end{Def} 

If $z \notin \Lambda_A$, then $\nu_z = 0$. Note that the index of the operator
$A$ itself is sometimes discussed~\cite{Atiy73}. In such contexts, the index of
$A$ is $\nu_0$. Hence, $\nu_\lambda$ corresponds to the index of $A - \lambda
I$.

The index of an eigenvalue gives information beyond what the algebraic and
geometric multiplicities themselves yield. Nevertheless, for $\lambda \in
\Lambda_A$, it is always true that $\nu_\lambda - 1 \leq a_\lambda - g_\lambda
\leq a_\lambda - 1$. In the diagonalizable case, $a_\lambda = g_\lambda$ and
$\nu_\lambda = 1$ for all $\lambda \in \Lambda_A$.

The following employs basic features of complex analysis extensively in
conjunction with linear algebra. Let us therefore review several elementary
notions in complex analysis. Recall that a \emph{holomorphic function} is one
that is complex differentiable throughout the domain under consideration. A
\emph{pole} of order $n$ at $z_0$ is a singularity that behaves as $h(z)/
(z-z_0)^n$ as $z \to z_0$, where $h(z)$ is holomorphic within a neighborhood of
$z_0$ and $h(z_0) \neq 0$.  We say that $h(z)$ has a \emph{zero} of order $m$
at $z_1$ if $1 / h(z)$ has a pole of order $m$ at $z_1$. A \emph{meromorphic
function} is one that is holomorphic except possibly at a set of isolated poles
within the domain under consideration.

Defined over the continuous complex variable $\z \in \mathbb{C}$, $A$'s
\emph{resolvent}:
\begin{align*}
\R(\z; \opGen) \equiv (z I - \opGen)^{-1} ~,
\end{align*}
captures all of $\opGen$'s spectral information through the poles of $\R(\z;
\opGen)$'s matrix elements. In fact, the resolvent contains more than just
$\opGen$'s spectrum: we later show that the order of each pole gives the index
$\nu$ of the corresponding eigenvalue. 

The spectrum $\Lambda_\opGen$ can be expressed in terms of the resolvent.
Explicitly, the \emph{point spectrum} (i.e., the set of eigenvalues) is the set of
complex values $z$ at which $zI-\opGen$ is not a one-to-one mapping,
with the implication that the inverse of $zI-\opGen$ does not exist:
\begin{align*}
\Lambda_\opGen = \bigl\{ \lambda \in \mathbb{C}:
  R(\lambda; \opGen)  \neq  \text{inv}(\lambda I - \opGen) \bigr\}
  ~,
\end{align*}
where $\text{inv}(\cdot)$ is the inverse of its argument. Later, via our
investigation of the Drazin inverse, it should become clear that the resolvent
operator can be self-consistently defined at the spectrum, despite the lack of
inverse.

For infinite-rank operators, the spectrum becomes more complicated.  In that
case, the right point spectrum (the point spectrum of $A$) need not be the same
as the left point spectrum (the point spectrum of $A$'s dual $A^\top$).
Moreover, the spectrum may grow to include non-eigenvalues $z$ for which the
range of $zI - A$ is not dense in the vector space it transforms or for which
$zI - A$ has dense range but the inverse of $zI - A$ is not bounded. These two
settings give rise to the so-called residual spectrum and continuous spectrum,
respectively~\cite{Kubr12}. To mitigate confusion, it should be noted that the
point spectrum can be continuous, yet never coincides with the continuous
spectrum just described. Moreover, understanding only countable point spectra
is necessary to follow the developments here.

Each of $A$'s eigenvalues $\lambda$ has an associated \emph{projection operator}
$A_\lambda$, which is the \emph{residue} of the resolvent as $z \to \lambda$~\cite{Hass99a}.
Explicitly:
\begin{align*}
A_\lambda = \text{Res} \bigl( (zI - A)^{-1} , \, z \to \lambda  \bigr) ~,
\end{align*}
where Res$( \, \cdot \, , z \to \lambda)$ is the element-wise residue of its first argument as $z \to \lambda$.
The projection operators are orthonormal:
\begin{align}
\label{eq:ProjOpsAreOrthonormal}
\opGen_\lambda \opGen_\zeta = \delta_{\lambda, \zeta} \opGen_\lambda
  ~. 
\end{align} 
and sum to the identity: 
\begin{align}
I  & = \sum_{\lambda \in \Lambda_\opGen} \opGen_{\lambda}
  ~. 
\label{eq:ProjOpsSumToIdentity}
\end{align} 
The following discusses in detail and then derives several new properties
of projection operators.

\section{Functional Calculi}
\label{sec:FunctionalCalculi}

In the following, we develop an extended \emph{functional calculus} that makes
sense of arbitrary functions $f(\cdot)$ of a linear operator $A$. Within any
functional calculus, one considers how $A$'s eigenvalues map to the eigenvalues
of $f(A)$; which we call a \emph{spectral mapping}. For example, it is known
that holomorphic functions of bounded linear operators enjoy an especially
simple spectral mapping theorem~\cite{Haas05}:
\begin{align*}
\Lambda_{f(A)} = f(\Lambda_A) ~.
\end{align*}
To fully appreciate the meromorphic functional calculus, we first state and
compare the main features and limitations of alternative functional calculi.

\subsection{Taylor series}

Inspired by the Taylor expansion of scalar functions: 
\begin{align*}
f(a) = \sum_{n=0}^{\infty} \frac{f^{(n)}(\xi) }{n!} \, (a - \xi)^n
  ~,
\end{align*}
a calculus for functions of an operator $A$ can be based on the series:
\begin{align}
    f(A) = \sum_{n=0}^{\infty} \frac{f^{(n)}(\xi) }{n!} \, (A - \xi I)^n ~,
\label{eq:TaylorFnlCalc}
\end{align}
where $f^{(n)}(\xi)$ is the $n^\text{th}$ derivative of $f(z)$ evaluated at $z=\xi$.

This is often used, for example, to express the exponential of $A$ as:
\begin{align*}
e^A = \sum_{n=0}^{\infty} \frac{A^{n} }{n!}
  ~.
\end{align*}
This particular series-expansion is convergent for any $A$ since $e^z$ is \emph{entire},
in the sense of complex analysis. Unfortunately, even if it exists there is a
limited domain of convergence for most functions. For example, suppose $f(z)$
has poles and choose a Maclaurin series; i.e., $\xi = 0$ in
Eq.~\eqref{eq:TaylorFnlCalc}. Then the series only converges when $A$'s
spectral radius is less than the radius of the innermost pole of $f(z)$.
Addressing this and related issues leads directly to alternative functional
calculi.

\subsection{Holomorphic functional calculus}

Holomorphic functions are well behaved, smooth functions that are complex
differentiable. Given a function $f(\cdot)$ that is holomorphic within a disk
enclosed by a counterclockwise contour $C$, its Cauchy integral formula is
given by:
\begin{align}
f(a) = \frac{1}{2 \pi i} \oint_C f(z) \, (z-a)^{-1} \, dz
  ~,
\label{eq:CauchyIForm}
\end{align}
Taking this as inspiration, the holomorphic functional calculus performs
a contour integration of the resolvent to extend $f(\cdot)$ to operators:
\begin{align}
f(A) = \frac{1}{2 \pi i} \oint_{C_{\Lambda_A}} f(z) \, (zI - A)^{-1} \, dz
  ~,
 \label{eq:HFCwResolvent}
\end{align}
where $C_{\Lambda_A}$ is a closed counterclockwise contour that encompasses
$\Lambda_A$. Assuming that $f(z)$ is holomorphic at $z = \lambda$ for all
$\lambda \in \Lambda_A$, a nontrivial calculation \cite{Dunf43a} shows that
Eq.~\eqref{eq:HFCwResolvent} is equivalent to the holomorphic calculus defined
by:
\begin{align}
 f(A) &= \sum_{\lambda \in \Lambda_A} \sum_{m=0}^{\nu_\lambda - 1} \frac{f^{(m)}(\lambda)}{m!} (A - \lambda I)^m A_\lambda
  ~.
\label{eq:HolomorphicFnlCalc}
\end{align}
After some necessary development, we will later derive Eq.~\eqref{eq:HolomorphicFnlCalc} as a special case of our meromorphic functional calculus, such that Eq.~\eqref{eq:HolomorphicFnlCalc} is valid whenever
$f(z)$ is holomorphic at $z = \lambda$ for all $\lambda \in \Lambda_A$.

The holomorphic functional calculus was first proposed in Ref. \cite{Dunf43a}
and is now in wide use; e.g., see Ref. \cite[p. 603]{Meyer00}. It agrees with
the Taylor-series approach whenever the infinite series converges, but gives an
functional calculus when the series approach fails. For example, using the
principal branch of the complex logarithm, the holomorphic functional calculus
admits $\log(A)$ for any nonsingular matrix, with the satisfying result that
$e^{\log(A)} = A$. Whereas, the Taylor series approach fails to converge for
the logarithm of most matrices even if the expansion for, say, $\log(1-z)$ is
used.

The major shortcoming of the holomorphic functional calculus is that it assumes
$f(z)$ is holomorphic at $\Lambda_A$. Clearly, if $f(z)$ has a pole at some $z
\in \Lambda_A$, then Eq.~\eqref{eq:HolomorphicFnlCalc} fails. An example of
such a failure is the negative-one power of a singular operator, which we take
up later on.

Several efforts have been made to extend the holomorphic functional calculus.
For example, Refs.~\cite{Gind66a} and~\cite{Nagy79a} define a functional
calculus that extends the standard holomorphic functional calculus to include a
certain class of meromorphic functions that are nevertheless still required to
be \emph{holomorphic on the point spectrum} (i.e., on the eigenvalues) of the
operator. However, we are not aware of any previous work that introduces and
develops the consequences of a functional calculus for functions that are
meromorphic on the point spectrum---which we take up in the next few sections.  

\subsection{Meromorphic functional calculus}
\label{sec:MFC_introduced}

Meromorphic functions are holomorphic except at a set of isolated poles of the function.  The resolvent of a finite-dimensional operator is meromorphic, since it is holomorphic everywhere except for poles at the eigenvalues of the operator.
We will now also allow our function $f(z)$ to be meromorphic with possible poles that coincide with the poles of the resolvent.

Inspired again by the Cauchy integral formula of Eq. (\ref{eq:CauchyIForm}),
but removing the restriction to holomorphic functions, our meromorphic
functional calculus instead employs a partitioned contour integration of the
resolvent:
\begin{align*}
f(\opGen) = \sum_{\lambda \in \Lambda_\opGen} \frac{1}{2 \pi i} \oint_{C_\lambda} f(z) R(z; \opGen) \, dz
  ~,
\end{align*}
where $C_\lambda$ is a small counterclockwise contour around the eigenvalue
$\lambda$. This and a spectral decomposition of the resolvent (to be derived
later) extends the holomorphic calculus to a much wider domain, defining:
\begin{align}
f(A)  & = \sum_{\lambda \in \Lambda_\opGen} \sum_{m = 0}^{\nu_\lambda - 1}
      \opGen_\lambda \bigl( \opGen -  \lambda I \bigr)^m 
      \frac{1}{2 \pi i} \oint_{C_\lambda} \frac{f(z)}{(z - \lambda)^{m+1}} \, dz  ~.
\label{eq:MFC_introduced}      
\end{align}      
The contour is integrated using knowledge of $f(z)$ since meromorphic $f(z)$
can introduce poles and zeros at $\Lambda_\opGen$ that interact with the
resolvent's poles.

The meromorphic functional calculus agrees with the Taylor-series approach
whenever the series converges and agrees with the holomorphic functional
calculus whenever $f(z)$ is holomorphic at $\Lambda_A$. However, when both the
previous functional calculi fail, the meromorphic calculus extends the domain
of $f(A)$ to yield surprising, yet sensible answers. For example, we show that
within it, the negative-one power of a singular operator is the Drazin
inverse---an operator that effectively inverts everything that is invertible.

The major assumption of our meromorphic functional calculus is that the domain
of operators must have a spectrum that is at most countably infinite---e.g.,
$A$ can be any compact operator. A related limitation is that singularities of
$f(z)$ that coincide with $\Lambda_A$ must be isolated singularities.
Nevertheless, we expect that these restrictions can be lifted with proper
treatment, as discussed in fuller context later.

\section{Meromorphic Spectral Decomposition}
\label{sec:EverythingDecomposed}

The preceding gave an overview of the relationship between alternative functional
calculi and their trade-offs, highlighting the advantages of the meromorphic
functional calculus. This section leverages these advantages and employs a
partial fraction expansion of the resolvent to give a general spectral
decomposition of almost any function of any operator. Then, since it plays a
key role in applications, we apply the functional calculus to investigate the
negative-one power of singular operators---thus \emph{deriving}, what is
otherwise an operator defined axiomatically, the Drazin inverse from first
principles.

\subsection{Partial fraction expansion of the resolvent} 

The elements of $\opGen$'s resolvent are proper rational functions that contain
all of $\opGen$'s spectral information. (Recall that a \emph{proper rational
function} $r(z)$ is a ratio of polynomials in $z$ whose numerator has degree
strictly less than the degree of the denominator.) In particular, the
resolvent's poles coincide with $\opGen$'s eigenvalues since, for $z \notin
\Lambda_\opGen$:
\begin{align}
\R(\z; \opGen) 
  & = (\z I - \opGen)^{-1} \nonumber \\
  & = \frac{\Cmatrix^\top}{\text{det}(\z I - \opGen)} \nonumber \\
  & = \frac{\Cmatrix^\top}
  {\prod_{\lambda \in \Lambda_\opGen} ( \z - \lambda )^{a_\lambda}}
  ~,  
\label{eq:Resolvent_as_CToverProduct}
\end{align}
where $a_\lambda$ is the algebraic multiplicity of eigenvalue $\lambda$ and
$\Cmatrix$ is the matrix of \emph{cofactors} of $\z I - \opGen$. That is,
$\Cmatrix$'s transpose $\Cmatrix^\top$ is the \emph{adjugate} of $\z I -
\opGen$:
\begin{align*}
\Cmatrix^\top = \text{adj}(\z I - \opGen)
  ~,
\end{align*}
whose elements will be polynomial functions of $z$ of degree less than $\sum_{\lambda \in \Lambda_A} a_\lambda$.

Recall that the partial fraction expansion of a proper rational function $r(z)$
with poles in $\Lambda$ allows a unique decomposition into a sum of constant
numerators divided by monomials in $z - \lambda$ up to degree $a_\lambda$, when
$a_\lambda$ is the order of the pole of $r(z)$ at $\lambda \in
\Lambda$~\cite{Latni98}. Equation (\ref{eq:Resolvent_as_CToverProduct}) thus
makes it clear that the resolvent has the unique partial fraction expansion:
\begin{align}
\R(\z; \opGen) 
  & = \sum_{\lambda \in \Lambda_\opGen} \sum_{m = 0}^{a_\lambda-1}
  \frac{1}{(\z - \lambda)^{m+1}}  A_{\lambda,m}
  ~,  
\label{eq:PartialFractionsExpansion_of_Resolvent_1}
\end{align} 
where $\{ A_{\lambda,m} \}$ is the set of matrices with constant entries
(\emph{not} functions of $\z$) uniquely determined elementwise by the partial fraction
expansion. However, $\R(\z; \opGen)$'s poles are \emph{not} necessarily of the
same order as the algebraic multiplicity of the corresponding eigenvalues since
the entries of $\Cmatrix$, and thus of $\Cmatrix^\top$, may have zeros at
$\opGen$'s eigenvalues. This has the potential to render $A_{\lambda,m}$ equal
to the zero matrix $\mathbf{0}$.

The Cauchy integral formula indicates that the constant matrices $\{
A_{\lambda,m} \}$ of Eq.~\eqref{eq:PartialFractionsExpansion_of_Resolvent_1}
can be obtained by the residues:
\begin{align}
A_{\lambda,m} = 
\frac{1}{2 \pi i} \oint_{C_{\lambda}} (\z - \lambda)^m \R(\z; \opGen)  d \z
~,
\label{eq:Alm}
\end{align}
where the residues are calculated elementwise. The projection operators
$A_\lambda$ associated with each eigenvalue $\lambda$ were already referenced
in \S \ref{sec:SpectralPrimer}, but can now be properly introduced as the
$A_{\lambda, 0}$ matrices:
\begin{align}
A_{\lambda} & = A_{\lambda, 0} \\
            & = \frac{1}{2 \pi i} \oint_{C_{\lambda}} \R(\z; \opGen)  d \z
  ~.
\label{eq:ProjOpsViaRes}
\end{align}

Since $\R(\z; \opGen)$'s elements are rational functions, as we just showed, it
is analytic except at a finite number of isolated singularities---at $\opGen$'s
eigenvalues. In light of the residue theorem, this motivates the
Cauchy-integral-like formula that serves as the starting point for the
meromorphic functional calculus:
\begin{align}
f(\opGen) & = \sum_{\lambda \in \Lambda_\opGen}
  \frac{1}{2 \pi i} \oint_{C_{\lambda}} f(\z) \R(\z; \opGen)  d \z 
  ~.
\label{eq:PartlyDecomposedCauchyIntegralFormula}
\end{align} 
Let's now consider several immediate consequences.

\subsection{Decomposing the identity}

Even the simplest applications of
Eq.~\eqref{eq:PartlyDecomposedCauchyIntegralFormula} yield insight. Consider
the identity as the operator function $f(A) = A^0 = I$ that corresponds to the
scalar function $f(z) = z^0 = 1$. Then,
Eq.~\eqref{eq:PartlyDecomposedCauchyIntegralFormula} implies:
\begin{align*}
I & = \sum_{\lambda \in \Lambda_\opGen}
  \frac{1}{2 \pi i} \oint_{C_\lambda} \R(\z; \opGen)  d \z \\
  & = \sum_{\lambda \in \Lambda_\opGen} \opGen_{\lambda} 
  ~.
\end{align*} 
This shows that the projection operators are, in fact, a decomposition of the
identity, as anticipated in Eq.~\eqref{eq:ProjOpsSumToIdentity}.

\subsection{Dunford decomposition, decomposed}

For $f(\opGen) = \opGen$, Eqs. \eqref{eq:PartlyDecomposedCauchyIntegralFormula}
and \eqref{eq:Alm} imply that:
\begin{align}
\opGen & = \sum_{\lambda \in \Lambda_\opGen} 
  \frac{1}{2 \pi i} \oint_{C_{\lambda}} \z \R(\z; \opGen)  d \z \nonumber \\
& = \sum_{\lambda \in \Lambda_\opGen} \! \!
  \left[ \lambda  
  \tfrac{1}{2 \pi i} \oint_{C_{\lambda}}  \! \! \R(\z; \opGen)  d \z  +
  \tfrac{1}{2 \pi i} \oint_{C_{\lambda}} \! (\z - \lambda) \R(\z; \opGen)  d \z
  \right] \nonumber \\  
  & = \sum_{\lambda \in \Lambda_\opGen} 
  \left(\lambda  A_{\lambda, 0} +   A_{\lambda, 1}  \right)
  ~.
\label{eq:DecompOfopGen}  
\end{align}  
We denote the important set of nilpotent matrices $A_{\lambda, 1}$ that project
onto the generalized eigenspaces by relabeling them:
\begin{align} 
  N_\lambda & \equiv \opGen_{\lambda, 1}  \\
  & = \frac{1}{2 \pi i} \oint_{C_{\lambda}} (\z - \lambda) \R(\z; \opGen)  d \z
  ~.
 \label{eq:NlambdaDef}
\end{align}

Equation~\eqref{eq:DecompOfopGen} is the unique \emph{Dunford
decomposition}~\cite{Dunf54a}: $\opGen = D + N$, where $D \equiv  \sum_{\lambda
\in \Lambda_\opGen} \lambda \opGen_{\lambda}$ is diagonalizable, $N \equiv
\sum_{\lambda \in \Lambda_\opGen}  N_{\lambda}$ is nilpotent, and $D$ and $N$
commute: $[D, N] = \mathbf{0}$. This is also known as the
\emph{Jordan--Chevalley decomposition}.

The special case where $\opGen$ is diagonalizable implies that $N = \mathbf{0}$. And so, Eq.~\eqref{eq:DecompOfopGen} simplifies to:
\begin{align*}
\opGen = \sum_{\lambda \in \Lambda_\opGen} {\lambda} \opGen_{\lambda}
  ~.
\end{align*}

\subsection{The resolvent, resolved}

As shown in Ref.~\cite{Hass99a} and can be derived from Eqs.~\eqref{eq:ProjOpsViaRes}
and \eqref{eq:NlambdaDef}:
\begin{align*}
\opGen_\lambda \opGen_\zeta & = \delta_{\lambda, \zeta} \opGen_\lambda
  \text{~and~} \\
  \opGen_\lambda N_\zeta & = \delta_{\lambda, \zeta} N_\lambda
  ~.
\end{align*}
Due to these, our spectral decomposition of the Dunford decomposition implies
that:
\begin{align} 
N_\lambda &
  = \opGen_\lambda \Bigl( \opGen - \sum_{\zeta \in \Lambda_\opGen} \zeta \opGen_\zeta \Bigr)  \nonumber \\
  &
  = \opGen_\lambda \bigl( \opGen -  \lambda \opGen_\lambda \bigr)  \nonumber \\
  & = \opGen_\lambda \bigl( \opGen -  \lambda I \bigr) 
  ~. 
\label{eq: Nilpotent operators in terms of projectors and W}
\end{align} 
Moreover:
\begin{align} 
\opGen_{\lambda, m} 
  & = \opGen_\lambda \bigl( \opGen -  \lambda I \bigr)^m 
  ~. 
\label{eq:ProjOpExpression4ResidueMatrices}
\end{align} 

It turns out that for $m > 0$: $\opGen_{\lambda, m} = N_\lambda^m$. (See also
Ref. \cite[p. 483]{Hass99a}.) This leads to a generalization of the projection
operator orthonormality relations of Eq.~\eqref{eq:ProjOpsAreOrthonormal}.
Most generally, the operators of $\{ A_{\lambda, m} \}$ are mutually related by:
\begin{align}
\opGen_{\lambda, m} \opGen_{\zeta, n} = \delta_{\lambda, \zeta} \opGen_{\lambda, m+n}
  ~. 
\label{eq:GenMatrixOrthogonalityRelation}
\end{align} 
Finally, if we recall that the index $\nu_\lambda$ is the dimension of the largest
associated subspace, we find that the index of $\lambda$ characterizes the
nilpotency of $N_\lambda$: $N_\lambda^m = \mathbf{0}$ for $m \geq \nu_\lambda$.
That is:
\begin{align}
\opGen_{\lambda, m} &= \mathbf{0} \qquad \text{for } m \geq \nu_\lambda
  ~. 
\end{align}

Returning to Eq. \eqref{eq:PartialFractionsExpansion_of_Resolvent_1}, we see
that all $A_{\lambda, m}$ with $m \geq \nu_\lambda$ are zero-matrices and so do
not contribute to the sum. Thus, we can rewrite Eq.
\eqref{eq:PartialFractionsExpansion_of_Resolvent_1} as:
\begin{align} 
\R(\z; \opGen) & = 
  \sum_{\lambda \in \Lambda_\opGen} \sum_{m = 0}^{\nu_\lambda - 1}
  \frac{1}{(\z - \lambda)^{m+1}}  A_{\lambda,m}
  \label{eq:PartialFractionsExpansion_of_Resolvent_2}
  \\ 
\intertext{or:}
\R(\z; \opGen) & = 
\sum_{\lambda \in \Lambda_\opGen} \sum_{m = 0}^{\nu_\lambda - 1}
  \frac{1}{(\z - \lambda)^{m+1}}  \opGen_\lambda \bigl( \opGen -  \lambda I \bigr)^m
  ~,
\label{eq:ResolventAsFnOfSpectrumAndOperators}   
\end{align}
for $z \notin \Lambda_{\opGen}$.

The following sections sometimes use $\opGen_{\lambda, m}$ in place of
$\opGen_\lambda \bigl( \opGen - \lambda I \bigr)^m$. This is helpful both for
conciseness and when applying Eq.~\eqref{eq:GenMatrixOrthogonalityRelation}.
Nonetheless, the equality in Eq.~\eqref{eq:ProjOpExpression4ResidueMatrices} is
a useful one to keep in mind.

\subsection{Meromorphic functional calculus}

In light of Eq.~\eqref{eq:PartlyDecomposedCauchyIntegralFormula},
Eq.~\eqref{eq:PartialFractionsExpansion_of_Resolvent_2} together with
Eq.~\eqref{eq:ProjOpExpression4ResidueMatrices} allow us to express any
function of an operator simply and solely in terms of its spectrum (i.e., its
eigenvalues for the finite dimensional case), its projection operators, and itself: 
\begin{align}
f(A)  & = \sum_{\lambda \in \Lambda_\opGen} \sum_{m = 0}^{\nu_\lambda - 1}
      \opGen_{\lambda, m} \, 
      \frac{1}{2 \pi i} \oint_{C_\lambda} \frac{f(z)}{(z - \lambda)^{m+1}} \, dz  ~.
\label{eq:MFC}      
\end{align}    
In obtaining Eq.~\eqref{eq:MFC} we finally derived
Eq.~\eqref{eq:MFC_introduced}, as promised earlier in
\S~\ref{sec:MFC_introduced}.  Effectively, by modulating the modes associated
with the resolvent's singularities, the scalar function $f(\cdot)$ is mapped to
the operator domain, where its action is expressed in each of $A$'s independent
subspaces.

\subsection{Evaluating the residues}

Interpretation aside, how does one use this result? Equation \eqref{eq:MFC}
says that the spectral decomposition of $f(\opGen)$ reduces to the evaluation
of several residues, where:
\begin{align*}
\text{Res} \bigl(  g(z) , \;  z \to\lambda  \bigr) = \frac{1}{2 \pi i} \oint_{C_\lambda} g(z) \, dz
  ~. 
\end{align*}
So, to make progress with Eq.\ \eqref{eq:MFC}, we must
evaluate function-dependent residues of the form:
\begin{align*}
\text{Res} \left( f(z) / (z - \lambda)^{m+1} , \, z \to \lambda \right)
  ~. 
\end{align*}

If $f(z)$ were holomorphic at each $\lambda$, then the order of the pole would
simply be the power of the denominator. We could then use Cauchy's differential
formula for holomorphic functions:
\begin{align}
f^{(n)}(a) &= \frac{n!}{2 \pi i} \oint_{C_a} \frac{f(z)}{(z-a)^{n+1}} dz 
  ~,
\label{eq:CauchyDiffFormula}
\end{align}
for $f(z)$ holomorphic at $a$. And, the meromorphic calculus would reduce to
the holomorphic calculus. Often, $f(z)$ will be holomorphic at least at
\emph{some} of $A$'s eigenvalues.  And so, Eq.~\eqref{eq:CauchyDiffFormula} is
still locally a useful simplification in those special cases.

In general, though, $f(z)$ introduces poles and zeros at $\lambda \in
\Lambda_A$ that change their orders. This is exactly the impetus for the
generalized functional calculus. The residue of a complex-valued function
$g(z)$ around its isolated pole $\lambda$ of order $n+1$ can be calculated from:
\begin{align*}
\text{Res} \bigl(  g(z) , \;  z \to\lambda  \bigr) 
  & = \frac{1}{n!} \,  \lim_{z \to \lambda} \, \frac{d^{n}}{{dz}^{n}}
  \left[  (z - \lambda)^{n+1}  g(z)  \right] 
  ~.
\end{align*} 

\subsection{Decomposing $\opGen^L$}

Equation \eqref{eq:MFC} says that we can explicitly derive the spectral decomposition of powers of the operator $\opGen$. 
Of course, we already did this for the special cases of $\opGen^0$ and
$\opGen^1$. The goal, though, is to do this in general.

For $f(\opGen) = \opGen^L \to f(z) = z^L$, $z=0$ can be either a zero or a pole
of $f(z)$, depending on the value of $L$. In either case, an eigenvalue of
$\lambda=0$ will distinguish itself in the residue calculation of $A^L$ via its
unique ability to change the order of the pole (or zero) at $z=0$. For example,
at this special value of $\lambda$ and for integer $L > 0$, $\lambda = 0$
induces poles that \emph{cancel} with the zeros of $f(z) = z^L$, since $z^L$
has a zero at $z=0$ of order $L$. For integer $L < 0$, an eigenvalue of
$\lambda = 0$ \emph{increases} the order of the $z=0$ pole of $f(z) = z^L$.
For all other eigenvalues, the residues will be as expected. Hence, 
from Eq.~\eqref{eq:MFC} and inserting $f(z) = z^L$, for any $L
\in \mathbb{C}$:
\begin{widetext}
\begin{align} 
\opGen^L 
  & = \Biggl[ \sum_{\lambda \in \Lambda_\opGen \atop \lambda \neq 0} \sum_{m = 0}^{\nu_\lambda - 1}
      \opGen_\lambda \bigl( \opGen -  \lambda I \bigr)^m \!\!\!
      \overbrace{ \left(  \frac{1}{2 \pi i} \oint_{C_\lambda} \frac{z^L}{(z - \lambda)^{m+1}} \, dz  \right)}^{
          =\frac{1}{m!}  \lim_{z \to \lambda} \, \frac{d^{m}}{{dz}^{m}} z^L  = 
	    \frac{ \lambda^{L-m}}{m!} \prod_{n=1}^m (L-n+1) } \Biggr] 
      + \left[ 0 \in \Lambda_\opGen \right] 
      \sum_{m = 0}^{\nu_0 - 1}
          \opGen_0  \opGen^m 
          \underbrace{ \left(  \frac{1}{2 \pi i} \oint_{C_0} z^{L-m-1} \, dz  \right)}_{= \delta_{L, m}  } \nonumber \\
    & = \Biggl[ \sum_{\lambda \in \Lambda_\opGen \atop \lambda \neq 0} \sum_{m = 0}^{\nu_\lambda - 1}
         \binom{L}{m} \lambda^{L-m}  
         \opGen_\lambda \bigl( \opGen -  \lambda I \bigr)^m \Biggr] 
         + \left[ 0 \in \Lambda_\opGen \right] 
          \sum_{m = 0}^{\nu_0 - 1}
          \delta_{L, m}  \opGen_0  \opGen^m 
          ~,
\label{eq: T^n generally}
\end{align}
\end{widetext}
where $\binom{L}{m}$ is the generalized binomial coefficient:
\begin{align}
\binom{L}{m} & = \frac{1}{m!} \prod_{n=1}^m (L-n+1) & \text{with } \binom{L}{0} = 1~,
\end{align}
and $[ 0 \in \Lambda_\opGen ]$ is the Iverson bracket which takes on value $1$
if zero is an eigenvalue of $\opGen$ and $0$ if not. 
$A_{\lambda, m}$ was replaced by $A_\lambda (A - \lambda I)^m$ 
to suggest the more explicit calculations involved with evaluating any $A^L$.
Equation \eqref{eq: T^n generally} applies to any linear operator with only isolated singularities in
its resolvent. 

If $L$ is a nonnegative integer such that $L \ge \nu_\lambda - 1$ for all
$\lambda \in \Lambda_\opGen$, then: 
\begin{align}
\opGen^L 
    & =   \sum_{\lambda \in \Lambda_\opGen \atop \lambda \neq 0} 
        \sum_{m=0}^{\nu_\lambda - 1}  \binom{L}{m} \lambda^{L-m}
		\opGen_{\lambda, m} 
  ~,
  \label{eq: W^L spectral decomp for positive integer L}		
\end{align}
where $\binom{L}{m}$ is now reduced
to the traditional binomial coefficient $L! / (m! (L-m)!)$.

\subsection{Drazin inverse}
\label{subsec:DrazinI}

If $L$ is any negative integer, then $\binom{-|L|}{m}$ can be written as a 
traditional binomial coefficient $(-1)^m \binom{\left| L \right| +m -1}{m}$, yielding:
\begin{align}
\opGen^{- \left| L \right|} 
  & = \sum_{\lambda \in \Lambda_\opGen \atop  \lambda \neq 0}
      \sum_{m=0}^{\nu_\lambda - 1}  
          (-1)^m \tbinom{\left| L \right| +m -1}{m}  \lambda^{- \left| L \right| - m}
		\opGen_{\lambda, m} 	
  ~,
\label{eq: W^L spectral decomp for negative integer L}		
\end{align}
for $- \left| L \right|  \in \{ -1, -2, -3, \dots \}$.

Thus, negative powers of an operator can be consistently defined even for
noninvertible operators. In light of Eqs. \eqref{eq: T^n generally} and
\eqref{eq: W^L spectral decomp for negative integer L}, it appears that the
zero eigenvalue does not even contribute to the function. It is well known, in
contrast, that it wreaks havoc on the naive, oft-quoted definition of a
matrix's negative power:
\begin{align*}
\opGen^{-1} \overset{?}{=} \frac{\text{adj}(\opGen)}{\text{det}(\opGen)}
  = \frac{\text{adj}(\opGen)}{\prod_{\lambda \in \Lambda_\opGen} \lambda^{a_\lambda}} ~,
\end{align*}
since this would imply dividing by zero.
If we can accept large positive powers of singular matrices---for which the
zero eigenvalue does not contribute---it seems fair to also accept negative
powers that likewise involve no contribution from the zero eigenvalue. 

Editorializing aside, we note that extending the definition of $\opGen^{-1}$ to the
domain including singular operators via Eqs.\ \eqref{eq: T^n generally} and
\eqref{eq: W^L spectral decomp for negative integer L} implies that: 
\begin{align*}
\opGen^{\left| L \right|} \opGen^{-\left| \ell \right|}
  & = \opGen^{-\left| \ell \right|} \opGen^{\left| L \right|} \\
  & = \opGen^{\left| L \right| -\left| \ell \right|} & \text{for } \left| L \right| \geq \left| \ell \right| + \nu_0
  ~,
\end{align*}
which is a very sensible and desirable condition.
Moreover, we find that
$\opGen \opGen^{-1} = I - \opGen_0$.

Specifically, the negative-one power of any square matrix is in general
\emph{not} the same as the matrix inverse since inv$(\opGen)$ need not exist.
However, it is consistently defined via Eq.\ \eqref{eq: W^L spectral decomp for
negative integer L} to be:
\begin{align}
\opGen^{-1}
  & = \sum_{\lambda \in \Lambda_\opGen \setminus \{0\} } 
		\sum_{m=0}^{\nu_\lambda - 1} 
		  (-1)^m \lambda^{- 1 - m}  \opGen_{\lambda, m} 
\label{eq:Neg1Power}
  ~. 	
\end{align}
This is the \emph{Drazin inverse} $A^\mathcal{D}$ of $\opGen$. 
Note that it is \emph{not} the same as the Moore--Penrose pseudo-inverse
\cite{Moor20a,Penr55a}.

Although the Drazin inverse is usually defined axiomatically to satisfy certain
criteria~\cite{BenIs03}, it is naturally \emph{derived} as the negative one
power of a singular operator in the meromorphic functional calculus. We can
check that it indeed satisfies the axiomatic criteria for the Drazin inverse,
enumerated according to historical precedent:
\begin{align*}
&(1^{\nu_0}) & A^{\nu_0} A^\mathcal{D} A = A^{\nu_0} \\
&(2) & A^\mathcal{D} A A^\mathcal{D} = A^\mathcal{D} \\
&(5) & [A, A^\mathcal{D}] = 0 ~,
\end{align*}
which gives rise to the Drazin inverse's moniker as the $\{ 1^{\nu_0}, 2,
5\}$-inverse~\cite{BenIs03}.

While $\opGen^{-1}$ always exists, the resolvent is nonanalytic at $z=0$ for a
singular matrix. Effectively, the meromorphic functional calculus removes the
nonanalyticity of the resolvent in evaluating $\opGen^{-1}$. As a result, 
as we can see from Eq.~\eqref{eq:Neg1Power}, the Drazin inverse
inverts what is invertible; the remainder is zeroed out.

Of course, whenever $\opGen$ \emph{is} invertible, $\opGen^{-1}$ is equal to
inv$(\opGen)$. However, we should not confuse this coincidence with
equivalence. Moreover, despite historic notation there is no reason that the
negative-one power should in general be equivalent to the inverse. Especially,
if an operator is not invertible! To avoid confusing $\opGen^{-1}$ with
inv$(\opGen)$, we use the notation $\opGen^{\mathcal{D}}$ for the Drazin
inverse of $\opGen$. Still, $\opGen^{\mathcal{D}} = \text{inv}(\opGen)$,
whenever $0 \notin \Lambda_{\opGen}$.

Amusingly, this extension of previous calculi lets us resolve an elementary but
fundamental question: What is $0^{-1}$?  It is certainly not infinity. Indeed,
it is just as close to negative infinity! Rather: $0^{-1} = 0 \neq
\text{inv}(0)$.

Although Eq.~\eqref{eq:Neg1Power} is a constructive way to build the Drazin
inverse, it imposes more work than is actually necessary. Using the meromorphic
functional calculus, we can derive a new, simple construction of the Drazin
inverse that requires only the original operator and the eigenvalue-0
projector.

First, assume that $\lambda$ is an isolated singularity of $\R(\z; \opGen)$
with finite separation at least $\epsilon$ distance from the nearest
neighboring singularity. And, consider the operator-valued function
$f_\lambda^\epsilon$ defined via the RHS of:
\begin{align*}
A_\lambda & = f_\lambda^\epsilon(A) \\
  & = \tfrac{1}{2 \pi i}
  \oint_{\lambda + \epsilon e^{i \phi}} (\zeta I - A)^{-1} \, d \zeta 
  ~,
\end{align*}
with $\lambda + \epsilon e^{i \phi}$ defining an $\epsilon$-radius circular
contour around $\lambda$. Then we see that:
\begin{align}
f_\lambda^\epsilon(z) 
&=  \tfrac{1}{2 \pi i} \oint_{\lambda + \epsilon e^{i \phi}} (\zeta  - z)^{-1} \, d \zeta  \nonumber \\
&= \bigl[ z \in \mathbb{C}: | z - \lambda | < \epsilon \bigr] ~, 
\label{eq:ScalarProjectionFunction}
\end{align}
where $[ z \in \mathbb{C}: | z - \lambda | < \epsilon ]$ is the Iverson bracket
that takes on value 1 if $z$ is within $\epsilon$-distance of $\lambda$ and 0
if not.

Second, we use this to find that, for any $c \in \mathbb{C} \setminus \{ 0 \}$:
\begin{align}
(A + cA_0)^{-1} 
& = \sum_{\lambda \in \Lambda_\opGen} \sum_{m = 0}^{\nu_\lambda - 1}
      \opGen_{\lambda, m} \, 
      \tfrac{1}{2 \pi i} \oint_{C_\lambda} \frac{\bigl( z + c f_0^\epsilon(z) \bigr)^{-1}}{(z - \lambda)^{m+1}} \, dz \nonumber \\
&= \opGen^\mathcal{D} + \sum_{m=0}^{\nu_0 - 1} \opGen_0 \opGen^m \tfrac{1}{2 \pi i} \oint_{C_0} \frac{(z+c)^{-1}}{z^{m+1}}  \nonumber \\
&= \opGen^\mathcal{D} + \sum_{m=0}^{\nu_0 - 1} \opGen_0 \opGen^m (-1)^m / c^{m+1}
  ~,
\label{eq:Drazin_from_cA0}
\end{align}
where we asserted that the contour $C_0$ exists within the finite
$\epsilon$-ball about the origin.

Third, we note that $A+cA_0$ is invertible for all $c \neq 0$; this can be
proven by multiplying each side of Eq.~\eqref{eq:Drazin_from_cA0} by $A +
cA_0$. Hence, $(A + cA_0)^{-1} = \text{inv}(A + cA_0)$ for all $c \neq 0$.

Finally, multiplying each side of Eq.~\eqref{eq:Drazin_from_cA0} by $I - A_0$,
and recalling that $A_{0,0} A_{0,m} = A_{0,m}$, we find a useful expression for
calculating the Drazin inverse of any linear operator $A$, given only $A$ and
$A_0$. Specifically:
\begin{align}
A^{\mathcal{D}} &= (I - A_0) (A + cA_0)^{-1} ~.
\label{eq:Drazin_from_simple_product_w_cA0}
\end{align}
which is valid for any $c \in \mathbb{C} \setminus \{ 0 \}$.
Eq.~\eqref{eq:Drazin_from_simple_product_w_cA0} generalizes the result found
specifically for $c=-1$ in Ref.~\cite{Roth76}.

For the special case of $c = -1$, it is worthwhile to also consider the
alternative construction of the Drazin inverse implied by
Eq.~\eqref{eq:Drazin_from_cA0}:
\begin{align}
\opGen^\mathcal{D} = (\opGen - \opGen_0)^{-1} + \opGen_0 \Bigl( \sum_{m=0}^{\nu_0 - 1} \opGen^m \Bigr) ~.
\end{align}
By a spectral mapping ($\lambda \to 1 - \lambda$, for $\lambda \in \Lambda_T$),
the Perron--Frobenius theorem and Eq.~\eqref{eq:Drazin_from_cA0} yield an
important consequence for any stochastic matrix $T$. The Perron--Frobenius
theorem guarantees that $T$'s eigenvalues along the unit circle are associated
with a diagonalizable subspace.  In particular, $\nu_1 = 1$. Spectral mapping
of this result means that $T$'s eigenvalue $1$ maps to the eigenvalue $0$ of
$I-T$ and $T_1 = (I-T)_0$. Moreover:
\begin{align*}
[(I-T) + T_1]^{-1} = (I-T)^\mathcal{D} + T_1 
  ~,
\end{align*}
since $\nu_0 = 1$. This corollary of Eq.~\eqref{eq:Drazin_from_cA0} (with $c =
1$) corresponds to a number of important and well known results in the theory
of Markov processes. Indeed, $Z \equiv (I-T+T_1)^{-1} $ is called the
\emph{fundamental matrix} in that setting~\cite{Keme60}.


\subsection{Consequences and generalizations}
\label{sec:GenFunctionalCalc}

For an infinite-rank operator $A$ with a continuous spectrum, the meromorphic
functional calculus has the natural generalization:
\begin{align}
f(A) = \frac{1}{2 \pi i} \oint_{C_{\Lambda_A}} f(z) (zI - A)^{-1} \, dz ~,
\label{eq:InfRankCalc}
\end{align}
where the contour $C_{\Lambda_A}$ encloses the (possibly continuous) spectrum
of $A$ without including any unbounded contributions from $f(z)$ outside of
$C_{\Lambda_A}$. The function $f(z)$ is expected to be meromorphic within
$C_{\Lambda_A}$. This again deviates from the holomorphic approach, since the
holomorphic functional calculus requires that $f(z)$ is analytic in a
neighborhood around the spectrum; see \S~VII of Ref.~\cite{Dunf67a}.
Moreover, Eq.~\eqref{eq:InfRankCalc} allows an extension of the functional
calculus of Refs.~\cite{Gind66a,Nagy79a,Berm99a}, since the function can be
meromorphic at the point spectrum in addition being meromorphic on the residual
and continuous spectra.

In either the finite- or infinite-rank case, whenever $f(z)$ \emph{is} analytic
in a neighborhood around the spectrum, the meromorphic functional calculus
agrees with the holomorphic. Whenever $f(z)$ is \emph{not} analytic in a
neighborhood around the spectrum, the function is undefined in the holomorphic
approach. In contrast, the meromorphic approach extends the function to the
operator-valued domain and does so with novel consequences.

In particular, when $f(z)$ is \emph{not analytic in a neighborhood around the
spectrum}---say $f(z)$ is nonanalytic within $A$'s spectrum at $\Xi_f \subset
\Lambda_A$---then we expect to lose both homomorphism and spectral mapping
properties:
\begin{itemize}
\setlength{\topsep}{0pt}
\setlength{\itemsep}{0pt}
\setlength{\parsep}{0pt}
\item Loss of homomorphism: $f_1(A) f_2(A) \neq (f_1 \circ f_2)(A)$;
\item Loss of naive spectral mapping:
	$f(\Lambda_A \setminus \Xi_f) \subset \Lambda_{f(A)}$.
\end{itemize}

A simple example of both losses arises with the Drazin inverse, above. There,
$f_1(z) = z^{-1}$. Taking this and $f_2(z) = z$ combined with singular operator
$A$ leads to the loss of homomorphism: $A^\mathcal{D} A \neq I$. As for the
second property, the spectral mapping can be altered for the candidate spectra
at $\Xi_f$ via pole--pole or pole--zero interactions in the complex contour
integral. For $f(A) = A^{-1}$, how does $A$'s eigenvalue of $0$ get mapped into
the new spectrum of $A^\mathcal{D}$?  A naive application of the spectral
mapping theorem might seem to yield an undefined quantity.  But, using the
meromorphic functional calculus self-consistently maps the eigenvalue as
$0^{-1} = 0$. It remains to be explored whether the full spectral mapping is
preserved for any function $f(A)$ under the meromorphic interpretation of
$f(\lambda)$.

It should now be apparent that extending functions via the meromorphic
functional calculus allows one to express novel mathematical properties, some
likely capable of describing new physical phenomena. At the same time, extra
care is necessary. The situation is reminiscent of the loss of commutativity in
non-Abelian operator algebra: not all of the old rules apply, but the gain in
nuance allows for mathematical description of important phenomena.

We chose to focus primarily on the finite-rank case here since it is
sufficient to demonstrate the utility of the general projection-operator
formalism. Indeed, there are ample nontrivial applications in the finite-rank
setting that deserve attention. To appreciate these, we now turn to address the
construction and properties of general eigenprojectors.

\section{Constructing Decompositions}
\label{sec:ConstructDecomp}

At this point, we see that projection operators are fundamental to functions of
an operator. This prompts the practical question of how to actually calculate
them. The next several sections address this by deriving expressions with both
theoretical and applied use. We first address the projection operators
associated with index-one eigenvalues. We then explicate the relationship
between eigenvectors, generalized eigenvectors, and projection operators for
normal, diagonalizable, and general matrices. Finally, we address how the
general results specialize in several common cases of interest. After these, we
turn to examples and applications.

\subsection{Projection operators of index-one eigenvalues}
\label{sec:IndexOneProjOps}

To obtain the projection operators associated with each index-one eigenvalue
$\lambda \in \{ \zeta \in \Lambda_\opGen: \nu_\zeta = 1 \}$, we apply the
meromorphic calculus to an appropriately chosen function of $\opGen$, finding:
\begin{align*}
\prod_{\zeta \in \Lambda_\opGen \atop \zeta \neq \lambda}
  (\opGen - \zeta I )^{\nu_\zeta} 
&=
  \sum_{\xi \in \Lambda_\opGen}  \sum_{m = 0}^{\nu_\xi - 1}
  \frac{\opGen_{\xi,m}}{2 \pi i} \bigointssss_{C_\xi} 
	\frac{ \prod_{\zeta \in \Lambda_\opGen \atop \zeta \neq \lambda}  (z - \zeta )^{\nu_\zeta}  }
	{(\z - \xi)^{m+1}} \, d\z 
  \\  
  & = \opGen_\lambda  
	\frac{1}{2 \pi i} \bigointssss_{C_\lambda} 
	\frac{ \prod_{\zeta \in \Lambda_\opGen \atop \zeta \neq \lambda}  (z - \zeta )^{\nu_\zeta}  }
	{z - \lambda} \, d\z 
 \\ 
  & = \opGen_\lambda  \prod_{\zeta \in \Lambda_\opGen \atop \zeta \neq \lambda}
  (\lambda - \zeta )^{\nu_\zeta} 
  ~. 
\end{align*}
Therefore, if $\nu_\lambda = 1$:
\begin{align} 
\opGen_\lambda & =
  \prod_{\zeta \in \Lambda_\opGen \atop \zeta \neq \lambda}
 	\left( \frac{\opGen - \zeta I }{\lambda - \zeta} \right)^{\nu_\zeta} 
  ~. 
\label{eq: proj operators for index-one eigs}
\end{align}
As convenience dictates in our computations, we let $\nu_\zeta \to a_\zeta -
g_\zeta + 1$ or even $\nu_\zeta \to a_\zeta$ in Eq.\ \eqref{eq: proj operators
for index-one eigs}, since multiplying $\opGen_\lambda$ by $(\opGen - \zeta
I) / (\lambda - \zeta)$ has no effect for $\zeta \in \Lambda_\opGen \setminus
\{ \lambda \}$ if $\nu_\lambda = 1$.

Equation~\eqref{eq: proj operators for index-one eigs} generalizes a well known
result that applies when the index of \emph{all} eigenvalues is one. That
is, when the operator is diagonalizable, we have:
\begin{align*}
\opGen_\lambda = \prod_{\zeta \in \Lambda_\opGen \atop \zeta \neq \lambda}
	\frac{\opGen - \zeta I }{\lambda - \zeta}
  ~.
\end{align*}
To the best of our knowledge, Eq.~\eqref{eq: proj operators for index-one
eigs} is original.

Since eigenvalues can have index larger than one, not all projection operators
of a nondiagonalizable operator can be found directly from Eq.~\eqref{eq: proj
operators for index-one eigs}. Even so, it serves three useful purposes.
First, it gives a practical reduction of the eigen-analysis by finding all
projection operators of index-one eigenvalues. Second, if there is only one
eigenvalue that has index larger than one---what we call the \emph{almost
diagonalizable case}---then Eq.~\eqref{eq: proj operators for index-one eigs},
together with the fact that the projection operators must sum to the identity,
\emph{does} give a full solution to the set of projection operators. Third,
Eq.~\eqref{eq: proj operators for index-one eigs} is a powerful theoretical
tool that we can use directly to spectrally decompose functions, for example,
of a stochastic matrix whose eigenvalues on the unit circle are guaranteed to
be index-one by the Perron--Frobenius theorem.

Although index-one expressions have some utility, we need a more general
procedure to obtain all projection operators of any linear operator. Recall
that, with full generality, projection operators can also be calculated
directly via residues, as in Eq.~\eqref{eq:ProjOpsViaRes}. 

An alternative procedure---one that extends a method familiar at least in quantum mechanics---is to obtain the projection operators via eigenvectors.  However, quantum mechanics always concerns itself with a subset of diagonalizable operators. What is the necessary generalization?
For one, left and right eigenvectors are no longer simply conjugate transposes
of each other. More severely, a full set of spanning eigenvectors is no longer
guaranteed and we must resort to \emph{generalized} eigenvectors. Since the
relationships among eigenvectors, generalized eigenvectors, and projection
operators are critical to the practical calculation of many physical
observables of complex systems, we collect these results in the next section.

\subsection{Eigenvectors, generalized eigenvectors, and projection operators}
\label{sec:ProjectorsAndEigenvectors}

Two common questions regarding projection operators are: Why not just use
eigenvectors? And, why not use the Jordan canonical form? First, the
eigenvectors of a defective matrix do not form a complete basis with which to
expand an arbitrary vector. One needs generalized eigenvectors for this.
Second, some functions of an operator require removing, or otherwise altering,
the contribution from select eigenspaces. This is most adroitly handled with
the projection operator formalism where different eigenspaces (correlates of
Jordan blocks) can effectively be treated separately. Moreover, even for simple
cases where eigenvectors suffice, the projection operator formalism simply can
be more calculationally or mathematically convenient.

That said, it is useful to understand the relationship between projection
operators and generalized eigenvectors. For example, it is often useful to
create projection operators from generalized eigenvectors. This section
clarifies their connection using the language of matrices. In the most general
case, we show that the projection operator formalism is usefully concise.

\subsubsection{Normal matrices}

Unitary, Hermitian, skew-Hermitian, orthogonal, symmetric, and skew-symmetric
matrices are all special cases of normal matrices. As noted, normal matrices
are those that commute with their Hermitian adjoint (complex-conjugate
transpose): $A A^{\dagger} = A^{\dagger} A$. Moreover, a matrix is normal if
and only if it can be diagonalized by a unitary transformation: $A = U \Lambda
U^{\dagger}$, where the columns of the unitary matrix $U$ are the orthonormal
right eigenvectors of $A$ corresponding to the eigenvalues ordered along the
diagonal matrix $\Lambda$. For an $M$-by-$M$ matrix $A$, the eigenvalues in
$\Lambda_A$ are ordered and enumerated according to the possibly degenerate
$M$-tuple $(\Lambda_A) = ( \lambda_1, \ldots, \lambda_M )$. Since an
eigenvalue $\lambda \in \Lambda_A$ has algebraic multiplicity $a_{\lambda} \ge
1$, $\lambda$ appears $a_{\lambda}$ times in the ordered tuple.

Assuming $\opGen$ is normal, each projection operator $A_{\lambda}$ can be
constructed as the sum of all ket--bra pairs of right-eigenvectors
corresponding to $\lambda$ composed with their conjugate transpose. We later
introduce bras and kets more generally via generalized eigenvectors of the
operator $A$ and its dual $A^\top$. However, since the complex-conjugate
transposition rule between dual spaces is only applicable to a ket basis
derived from a normal operator, we put off using the bra-ket notation for now
so as not to confuse the more familiar ``normal'' case with the general case.

To explicitly demonstrate this relationship between projection operators,
eigenvectors, and their Hermitian adjoints in the case of normality, observe
that:
\begin{align*}
A &= U \Lambda U^\dagger \\
& = 
\begin{bmatrix}
\vec{u}_1 & \vec{u}_2 & \cdots & \vec{u}_M
\end{bmatrix} 
\begin{bmatrix}
\lambda_1 & 0 & \cdots & 0 \\
0 & \lambda_2 & \cdots & 0 \\
\vdots & \vdots & \ddots & \vdots \\
0 & 0 & \cdots & \lambda_M
\end{bmatrix} 
\begin{bmatrix}
\vec{u}_1^\dagger \\ 
\vec{u}_2^\dagger \\ 
\vdots \\ 
\vec{u}_M^\dagger
\end{bmatrix} 
\\ 
& = 
\begin{bmatrix}
\lambda_1 \vec{u}_1 & \lambda_2 \vec{u}_2 & \cdots & \lambda_M \vec{u}_M
\end{bmatrix} 
\begin{bmatrix}
\vec{u}_1^\dagger \\ 
\vec{u}_2^\dagger \\ 
\vdots \\ 
\vec{u}_M^\dagger
\end{bmatrix} 
\\
& = 
\sum_{j=1}^{M}
\lambda_j \vec{u}_j \vec{u}_j^\dagger \\
& = \sum_{\lambda \in \Lambda_A} \lambda A_{\lambda} ~.
\end{align*}
Evidently, for normal matrices $A$: 
\begin{align*} 
A_{\lambda}
  = \sum_{j=1}^M \delta_{\lambda, \lambda_j} \vec{u}_j \vec{u}_j^\dagger
  ~.
\end{align*}
And, since $\vec{u}_i^\dagger \vec{u}_j = \delta_{i,j}$, we have an orthogonal
set $ \{ A_{\lambda} \}_{\lambda \in \Lambda_A}$ with the property that:
\begin{align*}
A_{\zeta} A_{\lambda}
& = \sum_{i=1}^M  \sum_{j=1}^M
  \delta_{\zeta, \lambda_i} \delta_{\lambda, \lambda_j} 
  \vec{u}_i \vec{u}_i^\dagger \vec{u}_j \vec{u}_j^\dagger \\
& = \sum_{i=1}^M  \sum_{j=1}^M
  \delta_{\zeta, \lambda_i} \delta_{\lambda, \lambda_j} 
  \vec{u}_i  \delta_{i,j} \vec{u}_j^\dagger \\
& = \sum_{i=1}^M
  \delta_{\zeta, \lambda_i} \delta_{\lambda, \lambda_i} 
  \vec{u}_i  \vec{u}_i^\dagger \\
& = \delta_{\zeta,\lambda} A_{\lambda}
~.
\end{align*}
Moreover: 
\begin{align*}
\sum_{\lambda \in \Lambda_A} A_{\lambda}
  & = \sum_{j=1}^M \vec{u}_j \vec{u}_j^\dagger \\
  & = U U^{\dagger} \\
  & = I  ~,
\end{align*}
and so on. All of the expected properties of projection operators can be
established again in this restricted setting.

The rows of $U^{-1} = U^{\dagger}$ are $\opGen$'s left-eigenvectors. In this
case, they are simply the conjugate transpose of the right-eigenvector. Note
that conjugate transposition is the familiar transformation rule between
ket and bra spaces in quantum mechanics (see e.g., Ref.~\cite{Saku11})---a consequence of the restriction to
normal operators, as we will show.  Importantly, a more general formulation of
quantum mechanics would \emph{not} have this same restricted correspondence
between the dual ket and bra spaces.

To elaborate on this point, recall that vector spaces admit dual spaces and
dual bases. However, there is no sense of a dual correspondence of a single ket
or bra without reference to a full basis~\cite{Halm58a}. Implicitly in quantum
mechanics, the basis is taken to be the basis of eigenstates of any Hermitian
operator, nominally since observables are self-adjoint.

To allude to an alternative, we note that $\vec{u}_j^\dagger \vec{u}_j$ is not
only the Hermitian form of inner product $\braket{ \vec{u}_j , \, \vec{u}_j }$ (where $\braket{\cdot, \cdot}$ denotes the inner product) of
the right eigenvector $\vec{u}_j$ with itself, but importantly also the simple
dot-product of the left eigenvector $\vec{u}_j^\dagger$ and the right
eigenvector $\vec{u}_j$, where $\vec{u}_j^\dagger$ acts as a linear functional
on $\vec{u}_j$. Contrary to the substantial effort devoted to the
inner-product-centric theory of Hilbert spaces, this latter interpretation of
$\vec{u}_j^\dagger \vec{u}_j$---in terms of linear functionals and a
left-eigenvector basis for linear functionals---is what generalizes to a
consistent and constructive framework for the spectral theory beyond normal
operators, as we will see shortly.

\subsubsection{Diagonalizable matrices}

By definition, diagonalizable matrices can be diagonalized, but not necessarily
via a unitary transformation. All diagonalizable matrices can nevertheless be
diagonalized via the transformation: $A = P \Lambda P^{-1}$, where the columns
of the square matrix $P$ are the not-necessarily-orthogonal right eigenvectors
of $A$ corresponding to the eigenvalues ordered along the diagonal matrix
$\Lambda$ and where the rows of $P^{-1}$ are $\opGen$'s left eigenvectors.
Importantly, the left eigenvectors need not be the Hermitian adjoint of the
right eigenvectors. As a particular example, this more general setting is
required for almost any transition dynamic of a Markov chain. In other words,
the transition dynamic of any interesting complex network with irreversible
processes serves as an example of a nonnormal operator.

Given the $M$-tuple of possibly-degenerate eigenvalues $(\Lambda_A) =
(\lambda_1, \, \lambda_2, \, \dots \, , \, \lambda_M )$, there is a
corresponding $M$-tuple of linearly-independent right-eigenvectors
$(\ket{\lambda_1}, \, \ket{\lambda_2}, \, \dots \, , \, \ket{\lambda_M})$ and a
corresponding $M$-tuple of linearly-independent left-eigenvectors
$(\bra{\lambda_1}, \, \bra{\lambda_2}, \, \dots \, , \, \bra{\lambda_M})$ such
that:
\begin{align*}
A \ket{\lambda_j} = \lambda_j \ket{\lambda_j} 
\end{align*}
and:
\begin{align*}
\bra{\lambda_j} A = \lambda_j \bra{\lambda_j} 
\end{align*}
with the orthonormality condition that:
\begin{align*}
\braket{\lambda_i | \lambda_j} = \delta_{i, j} ~.
\end{align*}
To avoid misinterpretation, we stress that the bras and kets that
appear above are the left and right eigenvectors, respectively,
and typically do \emph{not} correspond to complex-conjugate transposition.

With these definitions in place, the projection operators for a diagonalizable matrix can be written: 
\begin{align*} 
A_{\lambda} = \sum_{j=1}^M \delta_{\lambda, \lambda_j} \ket{\lambda_j} \bra{\lambda_j} ~. 
\end{align*}
Then:
\begin{align*} 
A & = \sum_{\lambda \in \Lambda_A}  \lambda A_{\lambda} \\
& = \sum_{j=1}^{M}
    \lambda_j  \ket{\lambda_j} \bra{\lambda_j} \\
& = 
\begin{bmatrix}
    \lambda_1 \ket{\lambda_1} & \lambda_2 \ket{\lambda_2} & \cdots & \lambda_M \ket{\lambda_M}
\end{bmatrix} 
\begin{bmatrix}
    \bra{\lambda_1} \\ 
    \bra{\lambda_2} \\ 
    \vdots \\ 
    \bra{\lambda_M}
\end{bmatrix} 
\\
& = 
\begin{bmatrix}
    \ket{\lambda_1} & \ket{\lambda_2} & \cdots & \ket{\lambda_M}
\end{bmatrix} 
\begin{bmatrix}
    \lambda_1 & 0 & \cdots & 0 \\
    0 & \lambda_2 & \cdots & 0 \\
    \vdots & \vdots & \ddots & \vdots \\
    0 & 0 & \cdots & \lambda_M
\end{bmatrix} 
\begin{bmatrix}
    \bra{\lambda_1} \\ 
    \bra{\lambda_2} \\ 
    \vdots \\ 
    \bra{\lambda_M}
\end{bmatrix} 
\\ 
& = 
P \Lambda P^{-1} ~.
\end{align*}
So, we see that the projection operators introduced earlier in a coordinate-free manner have a concrete representation in terms of left and right eigenvectors when the operator is diagonalizable.

\subsubsection{Any matrix}
\label{sec:ProjOpsFromGenEigvects}

Not all matrices can be diagonalized, but all square matrices can be put into
\emph{Jordan canonical form} via the transformation: $A = Y J
Y^{-1}$~\cite{Meyer00}. Here, the columns of the square matrix $Y$ are the
linearly independent right eigenvectors and generalized right eigenvectors
corresponding to the Jordan blocks ordered along the diagonal of the
block-diagonal matrix $J$. And, the rows of $Y^{-1}$ are the corresponding left
eigenvectors and generalized left eigenvectors, but reverse-ordered within each
block, as we will show.

Let there be $n$ Jordan blocks forming the $n$-tuple $(J_1, \, J_2, \, \dots
\, , \, J_n)$, with $1 \le n \le M$. The $k^{\text{th}}$ Jordan block $J_k$ has
dimension $m_k$-by-$m_k$:
\begin{align*}
J_k &= 
\underbrace{
\left. \,
\begin{bmatrix}
    \lambda_k & 1 & 0 & \cdots & 0 & 0 & 0 \\
    0 & \lambda_k & 1 &            &   & 0 & 0 \\
       & 0 & \lambda_k &   &   &   & 0 \\
    \vdots &  &  & \ddots & \ddots &  & \vdots \\
    0 &    &  &    & \lambda_k & 1 & 0 \\
    0 & 0 &  &    & 0 & \lambda_k & 1 \\
    0 & 0 & 0 & \cdots &    & 0 & \lambda_k
\end{bmatrix} 
\right\}
}_{m_k \text{ columns}}
& m_k \text{ rows}
\end{align*}
such that:
\begin{align*}
\sum_{k=1}^n m_k = M ~.
\end{align*} 

Note that eigenvalue $\lambda \in \Lambda_A$ corresponds to $g_\lambda$
different Jordan blocks, where $g_\lambda$ is the geometric multiplicity of the
eigenvalue $\lambda$. Indeed:
\begin{align*}
n = \sum_{\lambda \in \Lambda_A} g_\lambda ~.
\end{align*}
Moreover, the index $\nu_\lambda$ of the eigenvalue $\lambda$ is defined as the
size of the largest Jordan block corresponding to $\lambda$. So, we write this
in the current notation as: 
\begin{align*}
\nu_\lambda = \max \{ \delta_{\lambda, \lambda_k} m_k \}_{k=1}^n
  ~.
\end{align*}

If the index of any eigenvalue is greater than one, then the conventional
eigenvectors do not span the $M$-dimensional vector space. However, the set of
$M$ generalized eigenvectors does form a basis for the vector space~\cite{Axle97}.

Given the $n$-tuple of possibly-degenerate eigenvalues 
$(\Lambda_A) = (\lambda_1, \, \lambda_2, \, \dots \, , \, \lambda_n )$, 
there is a corresponding $n$-tuple of $m_k$-tuples of linearly-independent
generalized right-eigenvectors:
\begin{align*}
\left( ( \ket{\lambda_1^{(m)}} )_{m=1}^{m_1} , \, ( \ket{\lambda_2^{(m)}} )_{m=1}^{m_2}, \, \dots \, , \, ( \ket{\lambda_n^{(m)}} )_{m=1}^{m_n} \right)
  ~,
\end{align*}
where:
\begin{align*}
( \ket{\lambda_k^{(m)}} )_{m=1}^{m_k} \equiv \left( \ket{\lambda_k^{(1)}} , \, \ket{\lambda_k^{(2)}} , \, \dots \, , \,\ket{\lambda_k^{(m_k)}} \right) 
\end{align*}
and a corresponding $n$-tuple of $m_k$-tuples of linearly-independent
generalized left-eigenvectors:
\begin{align*}
\left( ( \bra{\lambda_1^{(m)}} )_{m=1}^{m_1} , \, ( \bra{\lambda_2^{(m)}} )_{m=1}^{m_2}, \, \dots \, , \, ( \bra{\lambda_n^{(m)}} )_{m=1}^{m_n} \right)
  ~,
\end{align*}
where:
\begin{align*}
( \bra{\lambda_k^{(m)}} )_{m=1}^{m_k} \equiv \left( \bra{\lambda_k^{(1)}} , \, \bra{\lambda_k^{(2)}} , \, \dots \, , \,\bra{\lambda_k^{(m_k)}} \right) 
\end{align*}
such that:
\begin{align}
(A - \lambda_k I ) \ket{\lambda_k^{(m+1)}} =  \ket{\lambda_k^{(m)}} 
\label{eq:RightGenRecursion}
\end{align}
and:
\begin{align}
\bra{\lambda_k^{(m+1)}} (A - \lambda_k I ) = \bra{\lambda_k^{(m)}}  
  ~,
\label{eq:LeftGenRecursion}
\end{align}
for $0 \leq m \leq m_k - 1$, 
where $\ket{\lambda_j^{(0)}} = \vec{0}$ and $\bra{\lambda_j^{(0)}} = \vec{0}$.
Specifically, $\ket{\lambda_k^{(1)}}$ and $\bra{\lambda_k^{(1)}}$ are 
conventional right and left eigenvectors, respectively.

Most directly, the generalized right and left eigenvectors can be found as the
nontrivial solutions to:
\begin{align*}
(A - \lambda_k I )^m \ket{\lambda_k^{(m)}} =  \vec{0} 
\end{align*}
and:
\begin{align*}
\bra{\lambda_k^{(m)}} (A - \lambda_k I )^m = \vec{0}
  ~,
\end{align*}
respectively.

It should be clear from Eq.~\eqref{eq:RightGenRecursion} and
Eq.~\eqref{eq:LeftGenRecursion} that:
\begin{align*}
\braket{\lambda_k^{(m)} | (A - \lambda_k I )^\ell | \lambda_k^{(n)}} 
  & = \braket{\lambda_k^{(m - \ell)} |  \lambda_k^{(n)}} \\
  & = \braket{\lambda_k^{(m)} | \lambda_k^{(n-\ell)}}
  ~,
\end{align*}
for $m, n,  \in \{ 0, 1, \dots, m_k \}$ and $\ell \geq 0$.
At the same time, it is then easy to show that: 
\begin{align*}
\braket{\lambda_k^{(m)} | \lambda_k^{(n)}} &= \braket{\lambda_k^{(m+n)} |
\lambda_k^{(0)}} = 0,  & \text{~if } m + n \leq m_k ~,
\end{align*}
where $m, n \in \{ 0, 1, \dots, m_k \}$.
Imposing appropriate normalization, we find that:
\begin{align}
\braket{\lambda_j^{(m)} | \lambda_k^{(n)}} = \delta_{j, k} \delta_{m + n, m_k + 1} ~.
\label{eq:GenEigenvectorOrthogonality}
\end{align}
Hence, we see that the left eigenvectors and generalized eigenvectors are a
dual basis to the right eigenvectors and generalized eigenvectors.
Interestingly though, within each Jordan subspace, \emph{the most generalized
left eigenvectors are dual to the least generalized right eigenvectors}, and
vice versa.

(To be clear, in this terminology ``least generalized'' eigenvectors are the
standard eigenvectors. For example, the $\bra{\lambda_k^{(1)}}$ satisfying the
standard eigenvector relation $\bra{\lambda_k^{(1)}} A = \lambda_k
\bra{\lambda_k^{(1)}}$ is the least generalized left eigenvector of subspace
$k$.  By way of comparison, the ``most generalized''' right eigenvector of
subspace $k$ is $\ket{\lambda_k^{(m_k)}}$ satisfying the most generalized
eigenvector relation $(A - \lambda_k I ) \ket{\lambda_k^{(m_k)}} =
\ket{\lambda_k^{(m_k - 1)}}$ for subspace $k$. The orthonormality relation
shows that the two are dual correspondents: $\braket{\lambda_k^{(1)} |
\lambda_k^{(m_k)}} = 1$, while all other eigen-bra--eigen-ket closures
utilizing these objects are null.)

With these details worked out, we find that 
the projection operators for a nondiagonalizable matrix can be written as: 
\begin{align} 
\opGen_{\lambda} = \sum_{k=1}^n \sum_{m = 1}^{m_k} 
  \delta_{\lambda, \lambda_k} \ket{\lambda_k^{(m)}} \bra{\lambda_k^{(m_k + 1 - m)}} ~.
\label{eq:ProjectorsViaGenEigenvectors}
\end{align}
And, we see that a projection operator includes all of its left and right
eigenvectors and all of its left and right generalized eigenvectors. This
implies that the identity operator must also have a decomposition in terms of
both eigenvectors and generalized eigenvectors:
\begin{align*} 
I & = \sum_{\lambda \in \Lambda_{\opGen}} \opGen_{\lambda} \\
  & = \sum_{k=1}^n \sum_{m = 1}^{m_k}
  \ket{\lambda_k^{(m)}} \bra{\lambda_k^{(m_k + 1 - m)}}
  ~. 
\end{align*}
Let $\bigl[ \ket{\lambda_k^{(m)}} \bigr]_{m=1}^{m_k}$ denote the column vector:
\begin{align*}
\bigl[ \ket{\lambda_k^{(m)}} \bigr]_{m=1}^{m_k} &= 
\begin{bmatrix}
 \ket{\lambda_k^{(1)}} \\
  \vdots \\
   \ket{\lambda_k^{(m_k)}}
\end{bmatrix} ~,
\end{align*}
and let $\bigl[ \bra{\lambda_k^{(m_k+1-m)}} \bigr]_{m=1}^{m_k}$ denote the column vector:
\begin{align*}
\bigl[ \bra{\lambda_k^{(m_k+1-m)}} \bigr]_{m=1}^{m_k} &= 
\begin{bmatrix}
 \bra{\lambda_k^{(m_k)}} \\
  \vdots \\
   \bra{\lambda_k^{(1)}}
\end{bmatrix} ~.
\end{align*}
Then, using the above results, 
and the fact that Eq.~\eqref{eq:LeftGenRecursion} implies that $\bra{\lambda_k^{(m+1)}} A  = \lambda_k \bra{\lambda_k^{(m+1)}} + \bra{\lambda_k^{(m)}}$,
we derive the explicit generalized-eigenvector decomposition of the nondiagonalizable operator $A$: 
\begin{align*} 
A & = \bigl( \sum_{\lambda \in \Lambda_A}  A_{\lambda} \bigr) A \\
  & = \sum_{k=1}^n \sum_{m = 1}^{m_k} 
     \ket{\lambda_k^{(m)}} \bra{\lambda_k^{(m_k + 1 - m)}} A \\
  & = \sum_{k=1}^n \sum_{m = 1}^{m_k} 
     \ket{\lambda_k^{(m)}} \left( \lambda_k \bra{\lambda_k^{(m_k + 1 - m)}} +  \bra{\lambda_k^{(m_k - m)}} \right)
	\\
\! \! & = 
\begin{bmatrix}
    \bigl[ \ket{\lambda_1^{(m)}} \bigr]_{m=1}^{m_1} \\ 
    \bigl[ \ket{\lambda_2^{(m)}} \bigr]_{m=1}^{m_2} \\ 
    \vdots \\ 
    \bigl[ \ket{\lambda_n^{(m)}} \bigr]_{m=1}^{m_n}
\end{bmatrix}^{\mathbf{\top}}  \! \!
\begin{bmatrix}
    J_1 & 0 & \cdots & 0 \\
    0 & J_2 & \cdots & 0 \\
    \vdots & \vdots & \ddots & \vdots \\
    0 & 0 & \cdots & J_n
\end{bmatrix} 
\begin{bmatrix}
    \bigl[ \bra{\lambda_1^{(m_1+1-m)}} \bigr]_{m=1}^{m_1} \\ 
    \bigl[ \bra{\lambda_2^{(m_2+1-m)}} \bigr]_{m=1}^{m_2} \\ 
    \vdots \\ 
    \bigl[ \bra{\lambda_n^{(m_n+1-m)}} \bigr]_{m=1}^{m_n}
\end{bmatrix} \\ 
  & = Y J Y^{-1}
  ~,
\end{align*}
where, defining $Y$ as:
\begin{align*}
Y = 
  \begin{bmatrix}
    \bigl[ \ket{\lambda_1^{(m)}} \bigr]_{m=1}^{m_1} \\ 
    \bigl[ \ket{\lambda_2^{(m)}} \bigr]_{m=1}^{m_2} \\ 
    \vdots \\ 
    \bigl[ \ket{\lambda_n^{(m)}} \bigr]_{m=1}^{m_n}
  \end{bmatrix}^{\mathbf{\top}} 
~,
\end{align*}
we are forced by Eq.~\eqref{eq:GenEigenvectorOrthogonality} to recognize that:
\begin{align*}
Y^{-1} = 
\begin{bmatrix}
    \bigl[ \bra{\lambda_1^{(m_1+1-m)}} \bigr]_{m=1}^{m_1} \\ 
    \bigl[ \bra{\lambda_2^{(m_2+1-m)}} \bigr]_{m=1}^{m_2} \\ 
    \vdots \\ 
    \bigl[ \bra{\lambda_n^{(m_n+1-m)}} \bigr]_{m=1}^{m_n}
\end{bmatrix} 
\end{align*}
since then $Y^{-1} Y = I$, and we recall that the inverse is guaranteed to be unique.

The above demonstrates an explicit construction for the Jordan canonical form.
One advantage we learn from this explicit decomposition is that the complete
set of left eigenvectors and left generalized eigenvectors (encapsulated in
$Y^{-1}$) can be obtained from the inverse of the matrix of the complete set of
right eigenvectors and generalized right eigenvectors (encoded in $Y$) and vice
versa. One unexpected lesson, though, is that the generalized left eigenvectors
appear in reverse order within each Jordan block.

Using Eqs.~\eqref{eq:ProjectorsViaGenEigenvectors} and
\eqref{eq:ProjOpExpression4ResidueMatrices} with
Eq.~\eqref{eq:LeftGenRecursion}, we see that the nilpotent operators
$A_{\lambda, m}$ with $m > 0$ further link the various generalized eigenvectors
within each subspace $k$.  Said more suggestively, generalized modes of a
nondiagonalizable subspace are necessarily cooperative.

It is worth noting that the left eigenvectors and generalized left eigenvectors
form a basis for all linear functionals of the vector space spanned by the
right eigenvectors and generalized right eigenvectors. Moreover, the left
eigenvectors and generalized left eigenvectors are exactly the dual basis to
the right eigenvectors and generalized right eigenvectors by their
orthonormality properties. However, neither the left nor right eigen-basis is a
priori more fundamental to the operator. Sympathetically, the right
eigenvectors and generalized eigenvectors form a (dual) basis for all linear
functionals of the vector space spanned by the left eigenvectors and
generalized eigenvectors.

\subsubsection{Simplified calculi for special cases}
\label{sec:MFC_in_special_cases}

In special cases, the meromorphic functional calculus reduces the general
expressions above to markedly simpler forms. And, this can greatly expedite
practical calculations and provide physical intuition. Here, we show which
reductions can be used under which assumptions.

For functions of operators with a countable spectrum, recall that the general form of the meromorphic functional calculus is:
\begin{align}
f(A)  & = \sum_{\lambda \in \Lambda_\opGen} \sum_{m = 0}^{\nu_\lambda - 1}
      \opGen_{\lambda, m} \, 
      \frac{1}{2 \pi i} \oint_{C_\lambda} \frac{f(z)}{(z - \lambda)^{m+1}} \, dz  ~.
\label{eq:MFC_again}      
\end{align}    
Equations~\eqref{eq:ProjOpExpression4ResidueMatrices} and
\eqref{eq:ProjectorsViaGenEigenvectors} gave the method to calculate
$A_{\lambda, m}$ in terms of eigenvectors and generalized eigenvectors.

When the operator is \emph{diagonalizable} (not necessarily normal), this
reduces to:
\begin{align}
f(A)  & = \sum_{\lambda \in \Lambda_\opGen} 
      \opGen_{\lambda} \, 
      \frac{1}{2 \pi i} \oint_{C_\lambda} \frac{f(z)}{(z - \lambda)} \, dz
	  ~,
\label{eq:MFC_diagonable}      
\end{align}
where $A_\lambda$ can now be constructed from conventional right and left
eigenvectors, although $\bra{\lambda_j}$ is \emph{not} necessarily the
conjugate transpose of $\ket{\lambda_j}$.

When the function is \emph{analytic} on the spectrum of the (not necessarily diagonalizable) operator, then our functional calculus reduces to the holomorphic functional calculus:
\begin{align}
f(A) & = \sum_{\lambda \in \Lambda_A}
  \sum_{m=0}^{\nu_\lambda - 1} \frac{f^{(m)}(\lambda)}{m!}  A_{\lambda,m}
  ~.
\label{eq:HFC_again}
\end{align}

When the function is \emph{analytic} on the spectrum of a \emph{diagonalizable}
(not necessarily normal) operator this reduces yet again to:
\begin{align}
    f(A) &= \sum_{\lambda \in \Lambda_A} f(\lambda) A_\lambda
	~.
\label{eq:HFC_diagonable}
\end{align}

When the function is \emph{analytic} on the spectrum of a \emph{diagonalizable}
(not necessarily normal) operator with \emph{no degeneracy} this reduces even
further to:
\begin{align}
\label{eq:HFC_diagonable_and_nondegenerate}
f(A) & = \sum_{\lambda \in \Lambda_A} f(\lambda)
  \frac{\ket{\lambda} \bra{\lambda} }{\braket{\lambda | \lambda}}
	~.
\end{align}

Finally, recall that an operator is \emph{normal} when it commutes with its
conjugate transpose. If the function is \emph{analytic} on the spectrum of a
\emph{normal} operator, then we recover the simple form enabled by the spectral
theorem of normal operators familiar in physics. That is,
Eq.~\eqref{eq:HFC_diagonable} is applicable, but now we have the extra
simplification that $\bra{\lambda_j}$ is simply the conjugate transpose of
$\ket{\lambda_j}$: $\bra{\lambda_j} = \ket{\lambda_j}^\dagger$.

\section{Examples and Applications}
\label{sec:Examples}

To illustrate the use and power of the meromorphic functional calculus, we now
adapt it to analyze a suite of applications from quite distinct domains. First,
we point to a set of example calculations for finite-dimensional operators of
stochastic processes. Second, we show that the familiar Poisson process is
intrinsically nondiagonalizable, and hint that nondiagonalizability may be
common more generally in semi-Markov processes. Third, we illustrate how
commonly the Drazin inverse arises in nonequilibrium thermodynamics, giving a
roadmap to developing closed-from expressions for a number of key observables.
Fourth, we turn to signal analysis and comment on power spectra of processes
generated by nondiagonalizable operators. Finally, we round out the
applications with a general discussion of Ruelle--Frobenius--Perron and Koopman
operators for nonlinear dynamical systems.

\subsection{Spectra of stochastic transition operators}
\label{sec:StochasticSpectra}

The preceding employed the notation that $A$ represents a general linear
operator. In the following examples, we reserve the symbol $T$ for the operator
of a stochastic transition dynamic. If the state-space is finite and has a
stationary distribution, then $T$ has a representation that is a nonnegative
row-stochastic---all rows sum to unity---transition matrix.

The transition matrix's nonnegativity guarantees that for each $\lambda \in
\Lambda_T$ its complex conjugate $\overline{\lambda}$ is also in $\Lambda_T$.
Moreover, the projection operator associated with the complex conjugate of
$\lambda$ is the complex conjugate of $T_\lambda$:
$\matHMM_{\overline{\lambda}} = \overline{\matHMM_{\lambda}}$.

If the dynamic induced by $T$ has a stationary distribution over the state
space, then the spectral radius of $T$ is unity and all of $T$'s eigenvalues
lie on or within the unit circle in the complex plane. The maximal eigenvalues
have unity magnitude and $1 \in \Lambda_T$. Moreover, an extension of the
Perron--Frobenius theorem guarantees that eigenvalues on the unit circle have
algebraic multiplicity equal to their geometric multiplicity. And, so,
$\nu_\zeta = 1$ for all $\zeta \in \{ \lambda \in \Lambda_T:  | \lambda | = 1
\}$.

$T$'s index-one eigenvalue of $\lambda=1$ is associated with stationarity of
the associated Markov process. $T$'s other eigenvalues on the unit circle are
roots of unity and correspond to deterministic periodicities within the process.

All of these results carry over from discrete to continuous time. In continuous
time, where $e^{t G} = T_{t_0 \to t_0 + t}$, $T$'s stationary eigenvalue of
unity maps to $G$'s stationary eigenvalue of zero. If the dynamic has a
stationary distribution over the state space, then the rate matrix $G$ is
row-sum zero rather than row-stochastic. $T$'s eigenvalues, on or within the
unit circle, map to $G$'s eigenvalues with nonpositive real part in the
left-hand side of the complex plane.

To reduce ambiguity in the presence of multiple operators, functions of
operators, and spectral mapping, we occasionally denote eigenvectors with
subscripted operators on the eigenvalues within the bra or ket. For example,
$\ket{0_G} = \ket{1_T} \neq \ket{0_\mathcal{G}} = \ket{1_\mathcal{T}} \neq
\ket{0_T}$ disambiguates the identification of $\ket{0}$ when we have operators
$G$, $T$, $\mathcal{G}$, and $\mathcal{T}$ with $T = e^{\tau G}$, $\mathcal{T}
= e^{\tau \mathcal{G}}$, and $0 \in \Lambda_G, \Lambda_{\mathcal{G}},
\Lambda_T$.

\subsection{Randomness and memory in correlated processes} 
\label{sec:FiniteD}

The generalized spectral theory developed here has recently been applied to
give the first closed-form expressions for many measures of complexity for
stochastic processes that can be generated by probabilistic finite
automata~\cite{Crut13a}. Rather than belabor the Kolmogorov--Chaitin notion of
complexity which is inherently uncomputable~\cite{Vita93a}, the new analytic
framework infuses \emph{computational mechanics}~\cite{Crut12a} with a means to
compute very practical answers about an observed system's organization and to
address the challenges of prediction.

For example, we can now answer the obvious questions regarding prediction: How
random is a process? How much information is shared between the past and the
future? How \emph{far} into the past must we look to predict what is
predictable about the future? How \emph{much} about the observed history must
be remembered to predict what is predictable about the future? And so on. The
Supplementary Materials of Ref.~\cite{Crut13a} exploit the meromorphic
functional calculus to answer these (and more) questions for the symbolic
dynamics of a chaotic map, the spacetime domain for an elementary cellular
automata, and the chaotic crystallographic structure of a close-packed
polytypic material as determined from experimental X-ray diffractograms.

In the context of the current exposition, the most notable feature of the
analyses across these many domains is that our imposed questions, which entail
tracking an observer's state of knowledge about a process, necessarily
\emph{induce} a nondiagonalizable metadynamic that becomes the central object
of analysis in each case. (This metadynamic is the so-called \emph{mixed-state
presentation} of Refs. \cite{Crut08a,Crut08b}.)

This theme, and the inherent nondiagonalizability of prediction, is explored in
greater depth elsewhere~\cite{Riec_forthcoming2}. We also found that another
nondiagonalizable dynamic is induced even in the context of quantum
communication when determining how much memory reduction can be achieved if we
generate a classical stochastic process using quantum mechanics~\cite{Riec16a}.

We mention the above nondiagonalizable metadynamics primarily as a pointer to
concrete worked-out examples where the meromorphic functional calculus has been
employed to analyze finitary hidden Markov processes via explicitly calculated,
generalized eigenvectors and projection operators. We now return to a more
self-contained discussion, where we show that nondiagonalizability can be
induced by the simple act of counting. Moreover, the theory developed is then
applied to deliver quick and powerful results.

\subsection{Poisson point processes}
\label{sec:Poisson}

The meromorphic functional calculus leads naturally to a novel perspective on
the familiar Poisson counting process---a familiar stochastic process class
used widely across physics and other quantitative sciences to describe
``completely random'' event durations that occur over a continuous domain
\cite{Barb91a,Smit58a,Gers02a,Beic06a}. The calculus shows that the basic
Poisson distribution arises as the signature of a simple nondiagonalizable
dynamic. More to the point, we derive the Poisson distribution directly,
without requiring the limit of the discrete-time binomial distribution, as
conventionally done~\cite{Boas66}.

\begin{figure}
\begin{center}
\includegraphics[width=0.45\textwidth]{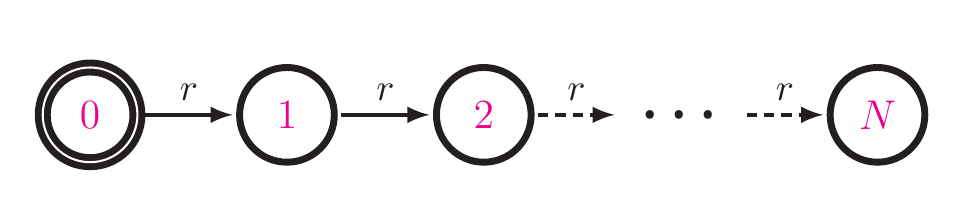}
\end{center}
\caption{Explicit Markov-chain representation of the continuous-time
  truncated Poisson dynamic, giving interstate transition rates $r$ among the
  first $N+1$ counter-states. (State self-transition rates $-r$ are not
  depicted.) Taking the limit of $N \to \infty$ recovers the full Poisson
  counting distribution. It can either be time-homogeneous (transition-rate
  parameter $r$ is time-independent) or time-inhomogeneous (parameter $r$ is
  time-dependent).
  }
\label{fig:PoissonCountingAutomata}
\end{figure}

Consider all possible counts, up to some arbitrarily large integer $N$. The
dynamics among these first $N+1$ counter states constitute what can be called
the \emph{truncated Poisson dynamic}. We recover the full Poisson distribution
as $N \to \infty$. A Markov chain for the truncated Poisson dynamic is shown in
Fig.~\ref{fig:PoissonCountingAutomata}. The corresponding rate matrix $G$, for
any arbitrarily large truncation $N$ of the possible count, is:
\begin{align*}
G = 
\begin{bmatrix}
-r & r  &   &      &    \\
   & -r & r &      &    \\
   &    &   \ddots & \ddots    & \\
   &    &    &   -r &  r  \\
   &    &    &      &  -r  \\
\end{bmatrix}
  ~,
\end{align*}
where $G_{ij}$ is the rate of transitioning to state (count) $j$ given that the
system is in state (count) $i$. Elements not on either the main diagonal or
first superdiagonal are zero. This can be rewritten succinctly as:
\begin{align*}
G = 
-r I + r D_1 ~,
\end{align*}
where $I$ is the identity operator in $N$-dimensions and $D_1$ is the
upshift-by-1 matrix in $N$-dimensions, with zeros everywhere, except $1$s along
the first superdiagonal. Let us also define the upshift-by-$m$ matrix $D_m$
with zeros everywhere except $1$s along the $m^\text{th}$ superdiagonal, such
that $D_m = D_1^m$ and $D_m^n = D_{m \cdot n}$, with $D_0 = I$. Operationally,
if $\bra{\delta_\ell}$ is the probability distribution over counter states that
is peaked solely at state $\ell$, then $\bra{\delta_\ell} D_m =
\bra{\delta_{\ell+m}}$.

For any arbitrarily large $N$, $G$'s eigenvalues are given by $\det (G- \lambda
I) = (-r-\lambda)^{N+1} = 0$, from which we see that its spectrum is the
singleton: $\Lambda_G = \{ -r \}$. Moreover, since it has algebraic
multiplicity $a_{-r} = N+1$ and geometric multiplicity $g_{-r} = 1$, the index
of the $-r$ eigenvalue is $\nu_{-r} = N+1$. Since $-r$ is the only eigenvalue,
and all projection operators must sum to the identity, we must have the
eigenprojection: $G_{-r} = I$. The lesson is that the Poisson point process is
highly nondiagonalizable.

\subsubsection{Homogeneous Poisson processes}

When the transition rate $r$ between counter states is constant in time, the net counter state-to-state transition operator from initial time $0$ to later time $t$ is given simply by:
\begin{align*}
T(t) = e^{t G} ~.
\end{align*}

The functional calculus allows us to directly evaluate $e^{t G}$ for the
Poisson nondiagonalizable transition-rate operator $G$; we find:
\begin{align*}
T(t) & = e^{t G} \\
     & = \sum_{\lambda \in \Lambda_G} \sum_{m=0}^{\nu_\lambda - 1}
	 G_\lambda (G - \lambda I)^m 
     \Bigl( \tfrac{1}{2 \pi i} \oint_{C_\lambda} \frac{e^{tz}}{(z - \lambda)^m} \, dz \Bigr) \\
     & = \lim_{N \to \infty} \sum_{m=0}^{N}
	 I (G + rI)^m \frac{1}{m!} \underbrace{\lim_{z \to -r} \frac{d^m}{dz^m} e^{tz}}_{t^m e^{-rt}} \\
     & = \sum_{m=0}^\infty (r D_1)^m \frac{t^m e^{-rt}}{m!} \\
     & = \sum_{m=0}^\infty D_m \frac{(rt)^m e^{-rt}}{m!} 
~.
\end{align*}

Consider the orthonormality relation $\braket{\delta_i | \delta_j} =
\delta_{i,j}$ between counter states, where $\ket{\delta_j}$ is represented by
$0$s everywhere except for a $1$ at counter-state $j$. It effectively measures
the occupation probability of counter-state $j$. Employing the result for
$T(t)$, we find the simple consequence that:
\begin{align*}
\bra{\delta_0} T(t) \ket{\delta_{n}}
  & =  \frac{(rt)^n e^{-rt}}{n!} \\
  & = \bra{\delta_m} T(t) \ket{\delta_{m+n}}
  ~.
\end{align*}
That is, the probability that the counter is incremented by $n$ in a time interval $t$ is independent of the initial count and given by:
$(rt)^n e^{-rt} / n!$.

Let us emphasize that these steps derived the Poisson distribution directly,
rather than as the typical limit of the binomial distribution.  Our derivation
depended critically on spectral manipulations of a highly nondiagonalizable
operator.  Moreover, our result for the transition dynamic $T(t)$ allows a
direct analysis of how \emph{distributions} over counts evolve in time, as
would be necessary, say, in a Bayesian setting with unknown prior count.  This
type of calculus can immediately be applied to the analysis of more
sophisticated processes, for which we can generally expect nondiagonalizability
to play an important functional role.

\subsubsection{Inhomogeneous Poisson processes}

Let us now generalize to time-inhomogeneous Poisson processes, where the
transition rate $r$ between count events is instantaneously uniform, but varies in time as $r(t)$.
Conveniently, the associated rate matrices at different times commute with each
other. Specifically,
with $G_a = -aI + aD_1$ and $G_b = -bI + bD_1$, we see that:
\begin{align*}
[G_a, \, G_b] = 0
  ~.
\end{align*}
Therefore, the net counter state-to-state transition operator from time $t_0$ to time $t_f$ is given by:
\begin{align}
T_{t_0, t_f} & = e^{\int_{t_0}^{t_f} G(t) \, dt}
             \nonumber \\
             & = e^{\left( \int_{t_0}^{t_f} r(t) \, dt \right) (-I + D_1)}
			              \nonumber \\
             & = e^{\braket{r} (\Delta t) (-I + D_1)}
			              \nonumber \\
             & = e^{ (\Delta t) G_{\braket{r}}}
~, 
\label{eq:etG_w_average_r}
\end{align}
where $\Delta t = t_f - t_0$ is the time elapsed and:
\begin{align*}
\braket{r} = \tfrac{1}{\Delta t} \int_{t_0}^{t_f} r(t) \, dt
\end{align*}
is the average rate during that time. Given Eq.~\eqref{eq:etG_w_average_r}, the
functional calculus proceeds just as in the time-homogeneous case to give the
analogous net transition dynamic:
\begin{align*}
T_{t_0, t_f} 
   & = \sum_{m=0}^\infty D_m
   \frac{\bigl( \braket{r} \Delta t \bigr)^m e^{- \braket{r} \Delta t}}{m!} 
~.
\end{align*}
The probability that the count is incremented by $n$ during the time interval
$\Delta t$ follows directly:
\begin{align*}
\bra{\delta_m} T_{t_0, t_f} \ket{\delta_{m+n}}
  & = \frac{\bigl( \braket{r} \Delta t \bigr)^n e^{- \braket{r} \Delta t}}{n!} 
  ~.
\end{align*}

With relative ease, our calculus allowed us to derive an important result for
stochastic process theory that is nontrivial to derive by other means.  Perhaps
surprisingly, we see that the probability distribution over final counts
induced by any rate trajectory $r(t)$ is the same as if the transition rate
were held fixed at mean $\braket{r}$ throughout the duration. Moreover, we can
directly analyze the net evolution of distributions over counts using the
derived transition operator $T_{t_0, t_f}$.

Note that the nondiagonalizability of the Poisson dynamic is robust in a
physical sense. That is, even varying the rate parameter in time in an erratic
way, the inherent structure of counting imposes a fundamental nondiagonalizable
nature. That nondiagonalizability can be robust in a physical sense is
significant, since one might otherwise be tempted to argue that
nondiagonalizability is extremely fragile due to numerical perturbations
within any matrix representation of the operator. This is simply not the case
since such perturbations are physically forbidden. Rather,
this simple example challenges us with the fact that some processes, even those
familiar and widely used, are intrinsically nondiagonalizable. On the positive
side, it appears that spectral methods can now be applied to analyze them. And,
this will be particularly important in more complex, memoryful processes
\cite{Marz14b,Marz14e,Marz15a}, including the hidden semi-Markov processes
\cite{Barb91a} that are, roughly speaking, the cross-product of hidden
finite-state Markov chains and renewal processes.

\subsection{Stochastic thermodynamics}
\label{sec:NoneqThermo}

The previous simple examples started to demonstrate the methods of the
meromorphic functional calculus. Next, we show a novel application of the
meromorphic functional calculus to environmentally driven mesoscopic dynamical
systems, selected to give a new set of results within nonequilibrium
thermodynamics. In particular, we analyze functions of singular transition-rate
operators. Notably, we show that the Drazin inverse arises naturally in the general solution
of Green--Kubo relations. We mention that it also arises when analyzing moments of the excess
heat produced in the driven transitions atop either equilibrium steady states
or nonequilibrium steady states~\cite{Riec_forthcoming}.

\subsubsection{Dynamics in independent eigenspaces}

An important feature of the functional calculus is its ability to address
particular eigenspaces independently when necessary. This feature is often
taken for granted in the case of normal operators; say, in physical dynamical
systems when analyzing stationary distributions or dominant decay modes.
Consider a singular operator $\mathcal{L}$ that is not necessarily normal and
not necessarily diagonalizable and evaluate the simple yet ubiquitous integral
$\int_0^\tau e^{t \mathcal{L}} \, dt$. Via the meromorphic functional calculus
we find:
\begin{align}
\int_0^\tau e^{t \mathcal{L}} \, dt
  & = \sum_{\lambda \in \Lambda_{\mathcal{L}} }
  \sum_{m=0}^{\nu_\lambda - 1} \mathcal{L}_{\lambda, m}
  \tfrac{1}{2 \pi i} \oint_{C_\lambda}
  \frac{ \int_{0}^\tau e^{t z} \, dt}{(z - \lambda)^{m+1}} \, dz
  \nonumber \\
  & = \Bigl( \sum_{m=0}^{\nu_0 - 1} \mathcal{L}_{0,m} \tfrac{1}{2 \pi i}
  \oint_{C_0} \frac{z^{-1} (e^{\tau z} - 1)}{z^{m+1}} \, dz  \Bigr)
  \nonumber \\ 
  & \quad + \sum_{\lambda \in \Lambda_{\mathcal{L}} \setminus 0} 
  \sum_{m=0}^{\nu_\lambda - 1}
  \mathcal{L}_{\lambda, m} \tfrac{1}{2 \pi i}
  \oint_{C_\lambda} \frac{z^{-1} (e^{\tau z} - 1)}{(z - \lambda)^{m+1}} \, dz
  \nonumber \\
  & = \Bigl( \sum_{m=0}^{\nu_0 - 1}
  \tfrac{\tau^{m+1}}{(m+1)!} \mathcal{L}_{0,m}  \Bigr)
  + \mathcal{L}^\mathcal{D} \left( e^{\tau \mathcal{L} } - I \right)
  ~,
\label{eq:IntOfExp}
\end{align}
where $\mathcal{L}^\mathcal{D}$ is the Drazin inverse of $\mathcal{L}$, discussed earlier.

The pole--pole interaction ($z^{-1}$ with $z^{-m-1}$) at $z=0$
distinguished the $0$-eigenspace in the calculations
and required the meromorphic functional calculus for direct analysis.
The given solution to this integral will be useful in the following.

Next, we consider the case where $\mathcal{L}$ is the transition-rate operator
among the states of a structured stochastic dynamical system. This leads to
several novel consequence within stochastic thermodynamics.

\subsubsection{Green--Kubo relations}

Let us reconsider the above integral in the case when the singular operator 
$\mathcal{L}$---let us call it $G$---is a transition-rate operator that
exhibits a single stationary distribution. By the spectral mapping $\ln \Lambda_{e^{G}}$ of the
eigenvalue $1 \in \Lambda_{e^{G}}$ addressed in the Perron--Frobenius theorem, 
$G$'s zero eigenmode is diagonalizable. 
And, by assuming a single attracting stationary distribution, the zero eigenvalue has algebraic multiplicity $a_0 = 1$.
Equation~\eqref{eq:IntOfExp} then simplifies to:
\begin{align}
\int_0^\tau e^{t G} \, dt
  & = \tau \ket{0_{G}} \bra{0_{G}}
  + G^\mathcal{D} \left( e^{\tau G } - I \right)
  ~.
\label{eq:IntOfExpRate}
\end{align}
Since $G$ is a transition-rate operator, the above integral corresponds to
integrated time evolution. The Drazin inverse $G^\mathcal{D}$ concentrates on
the transient contribution beyond the persistent stationary background. In
Eq.~\eqref{eq:IntOfExpRate}, the subscript within the left and right
eigenvectors explicitly links the eigenvectors to the operator $G$, to reduce
ambiguity.  Specifically, the projector $\ket{0_G} \bra{0_G}$ maps any
distribution to the stationary distribution.

Green--Kubo-type relations~\cite{Green54, Zwan65} connect the
out-of-steady-state transport coefficients to the time integral of steady-state
autocorrelation functions. They are thus very useful for understanding
out-of-steady-state dissipation due to steady-state fluctuations. (Steady state
here refers to either equilibrium or nonequilibrium steady state.)
Specifically, the Green--Kubo relation for a transport coefficient, $\kappa$
say, is typically of the form:
\begin{align*}
\kappa = \int_{0}^\infty \bigl( \Braket{ A(0) A(t)}_{\text{s.s.}} - \braket{A}_{\text{s.s.}}^2 \bigr) \, dt
  ~,
\end{align*}
where $A(0)$ and $A(t)$ are some observable of the stationary stochastic
dynamical system at time $0$ and time $t$, respectively, and the subscript
$\braket{ \cdot }_{\text{s.s.}} $ emphasizes that the expectation value is to
be taken according to the steady-state distribution.

Using:
\begin{align*}
\Braket{ A(0) A(t)}_{\text{s.s.}}
  & = \text{tr} \bigl( \ket{0_G} \bra{0_G} A \, e^{t G} A \bigr) \\
  & = \bra{0_G} A \, e^{t G} A \ket{0_G}
  ~,
\end{align*}
the transport coefficient $\kappa$ can be written more explicitly in terms of
the relevant transition-rate operator $G$ for the stochastic dynamics:
\begin{align}
\kappa & = \lim_{\tau \to \infty}
  \int_{0}^\tau \bra{0_G} A \, e^{t G} A \ket{0_G} \, dt 
  - \tau \bra{0_G} A \ket{0_G}^2 \nonumber \\
  & = \lim_{\tau \to \infty} \bra{0_G} A
  \Bigl(  \int_{0}^\tau e^{t G} \, dt  \Bigr) A \ket{0_G} 
  - \tau \bra{0_G} A \ket{0_G}^2 \nonumber \\
  & = \lim_{\tau \to \infty}  \bra{0_G} A \, G^\mathcal{D}
  \bigl( e^{\tau G} - I \bigr) A \ket{0_G} \nonumber \\
  & = - \braket{A \, G^\mathcal{D} A}_{\text{s.s.}}
  ~.
\label{eq:GCDrazin}
\end{align}
Thus, we learn that relations of Green--Kubo form are direct signatures of
the Drazin inverse of the transition-rate operator for the stochastic dynamic.

The result of Eq.~\eqref{eq:GCDrazin} holds quite generally. For example, if
the steady state has some number of periodic flows, the result of
Eq.~\eqref{eq:GCDrazin} remains valid. Alternatively, in the case of
nonperiodic chaotic flows---where $G$ will be the logarithm of the
Ruelle--Frobenius--Perron operator, as described later in
\S~\ref{sec:Ruelle_and_Koopman_operators}---$\ket{0_G} \bra{0_G}$ still induces
the average over the steady-state trajectories.

In the special case where the transition-rate operator is diagonalizable,
$- \braket{A \, G^\mathcal{D} A}_{\text{s.s.}}$ is simply the integrated
contribution from a weighted sum of decaying exponentials. Transport
coefficients then have a solution of the simple form:
\begin{align}
\kappa & = - \! \! \sum_{\lambda \in \Lambda_G \setminus 0}
  \frac{1}{\lambda} \bra{0_G} A \, G_\lambda  A \ket{0_G}
  ~.
\label{eq:GCDrazinSimpleEigenExpansion}
\end{align}
Note that the minus sign keeps $\kappa$ positive since Re$(\lambda) < 0$ for
$\lambda \in \Lambda_G \setminus \{ 0 \}$. Also, recall that $G$'s eigenvalues
with nonzero imaginary part occur in complex-conjugate pairs and
$G_{\overline{\lambda}} = \overline{G_\lambda}$. Moreover, if $G_{i,j}$ is the
classical transition-rate from state $i$ to state $j$ (to disambiguate from the
transposed possibility), then $\bra{0_G}$ is the stationary distribution. (The
latter is sometimes denoted $\bra{\boldsymbol{\pi}}$ in the Markov process
literature.) And, $\ket{0_G}$ is a column vector of all ones (sometimes denoted
$\ket{\one}$) which acts to integrate contributions throughout the state space.

A relationship of the form of Eq.~\eqref{eq:GCDrazin}, between the Drazin
inverse of a classical transition-rate operator and a particular Green--Kubo
relation was recently found in Ref.~\cite{Mandal15} for the friction tensor for
smoothly-driven transitions atop nonequilibrium steady states. Subsequently, a
truncation of the eigen-expansion of the form of
Eq.~\eqref{eq:GCDrazinSimpleEigenExpansion} was recently used in a similar
context to bound a universal tradeoff between power, precision, and
speed~\cite{Lahi16}. Equation~\eqref{eq:GCDrazin} shows that a fundamental
relationship between a physical property and a Drazin inverse is to be
expected more generally whenever the property can be related to integrated
correlation.


Notably, if a Green--Kubo-like relation integrates a cross-correlation,
say between
$A(t)$ and $B(t)$ rather than an autocorrelation,
then we have only the slight modification:
\begin{align}
\int_{0}^\infty \bigl( \Braket{ A(0) B(t)}_{\text{s.s.}} \! - \braket{A}_{\text{s.s.}} \!\! \braket{B}_{\text{s.s.}} \bigr) \, dt 
= - \braket{A \, G^\mathcal{D} B}_{\text{s.s.}}
~.
\label{eq:genGKsoln}
\end{align}

The foregoing analysis bears on both classical and quantum dynamics.
$G$ may be a so-called linear \emph{superoperator} in the quantum
regime~\cite{Lowd82}; for example, the \emph{Lindblad
superoperator}~\cite{Lind76,Barn00} that operates on density operators.  A
Liouville-space representation~\cite{Petr97} of the superoperator, though,
exposes the superficiality of the distinction between superoperator and
operator. At an abstract level, time evolution can be discussed uniformly
across subfields and reinterpretations of Eq.~\eqref{eq:genGKsoln} will be found
in each associated physical theory.

Reference~\cite{Riec_forthcoming} presents additional constructive results that
emphasize the ubiquity of integrated correlation and Drazin inverses in the
transitions between steady states~\cite{Oono98}, relevant to the fluctuations
within any physical dynamic. Overall, these results support the broader notion
that dissipation depends on the structure of correlation.

Frequency-dependent generalizations of integrated correlation have a
corresponding general solution. To be slightly less abstract, later on we give
novel representative formulae for a particular application: the general
solution to power spectra of a process generated by any countable-state hidden
Markov chain.

\begin{figure}[t]
\begin{center}
\includegraphics[width=0.375\textwidth]{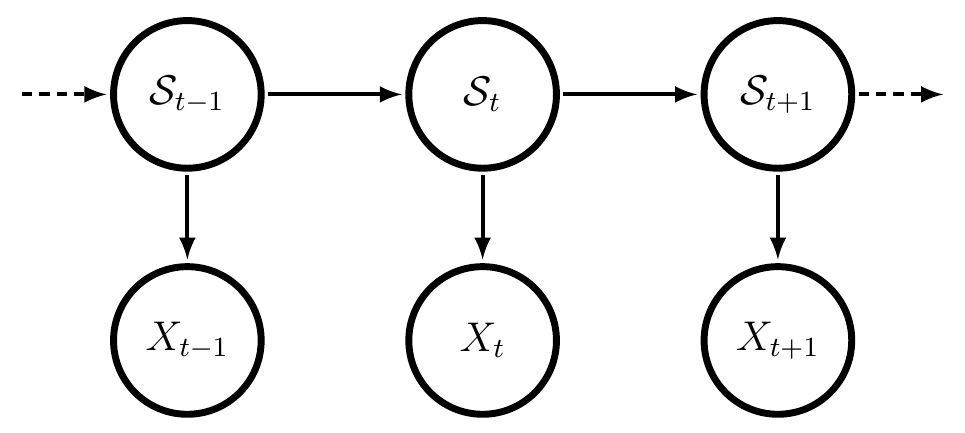}
\end{center}
\caption{Bayes network for a state-emitting hidden Markov model graphically
	depicts the structure of conditional independence among random variables
	for the latent state $\{ \St_n \}_{n \in \mathbb{Z}}$ at each time $n$ and
	the random variables $\{ X_n \}_{n \in \mathbb{Z}}$ for the observation at
	each time $n$.
  }
\label{fig:HMM_BayesNet}
\end{figure}

\subsection{Power spectra}
\label{sec:PSD}

A signal's power spectrum quantifies how its power is distributed across
frequency~\cite{Stoi05}. For a discrete-domain process it is:
\begin{align}
P(\omega) = \lim_{N \to \infty} \tfrac{1}{N} \Braket{ \Bigl| \sum_{n=1}^{N} X_n e^{-i \omega n} \Bigr|^2 } ~,
\label{eq:PSDdef}
\end{align}
where $\omega$ is the angular frequency and $X_n$ is the random variable for
the observation at time $n$. For a wide-sense stationary stochastic process, the power
spectrum is also determined from the signal's autocorrelation function
$\gamma(\tau)$:
\begin{align}
P(\omega) = \lim_{N \to \infty} \tfrac{1}{N} \sum_{\tau=-N}^{N} \bigl( N - \left| \tau \right| \bigr) \gamma(\tau) e^{-i \omega \tau}
  ~,
\label{eq:PSDfromACF}
\end{align}
where the autocorrelation function for a wide-sense stationary stochastic process is
defined:
\begin{align*}
\gamma(\tau) = \Braket{ \, \overline{X_n} X_{n + \tau}}_n
  ~.
\end{align*}
The windowing function $N - \left| \tau \right|$ appearing in
Eq.~\eqref{eq:PSDfromACF} is a direct consequence of Eq.~\eqref{eq:PSDdef}. It
is not imposed externally, as is common practice in signal analysis. This is
important to subsequent derivations.

The question we address is how to calculate the correlation function and power
spectrum given a model of the signal's generator. To this end, we briefly
introduce hidden Markov models as signal generators and then use the
meromorphic calculus to calculate their autocorrelation and power spectra in
closed-form. This leads to several lessons. First, we see that the power
spectrum is a direct fingerprint of the resolvent of the generator's
time-evolution operator, analyzed along the unit circle. Second, spectrally
decomposing the not-necessarily-diagonalizable time evolution operator, we
derive the most general qualitative behavior of the autocorrelation function
and power spectra. Third, contributions from eigenvalues on the unit circle
must be extracted and dealt with separately. Contributions from eigenvalues on
the unit circle correspond to Dirac delta functions---the analog of Bragg peaks
in diffraction. Whereas, eigen-contributions from inside the unit circle
correspond to diffuse peaks, which become sharper for eigenvalues closer to the
unit circle. Finally, nondiagonalizable eigenmodes yield qualitatively
different line profiles than their diagonalizable counterparts. In short, when
applied to signal analysis our generalized spectral decomposition has directly
measurable consequences. This has been key to analyzing low-dimensional
disordered materials, for example, when adapted to X-ray diffraction spectra
\cite{Star69,Riec14a,Riec14b}.

\begin{figure}
\begin{center}
\begin{overpic}[width=.375\textwidth,unit=1mm] 
      {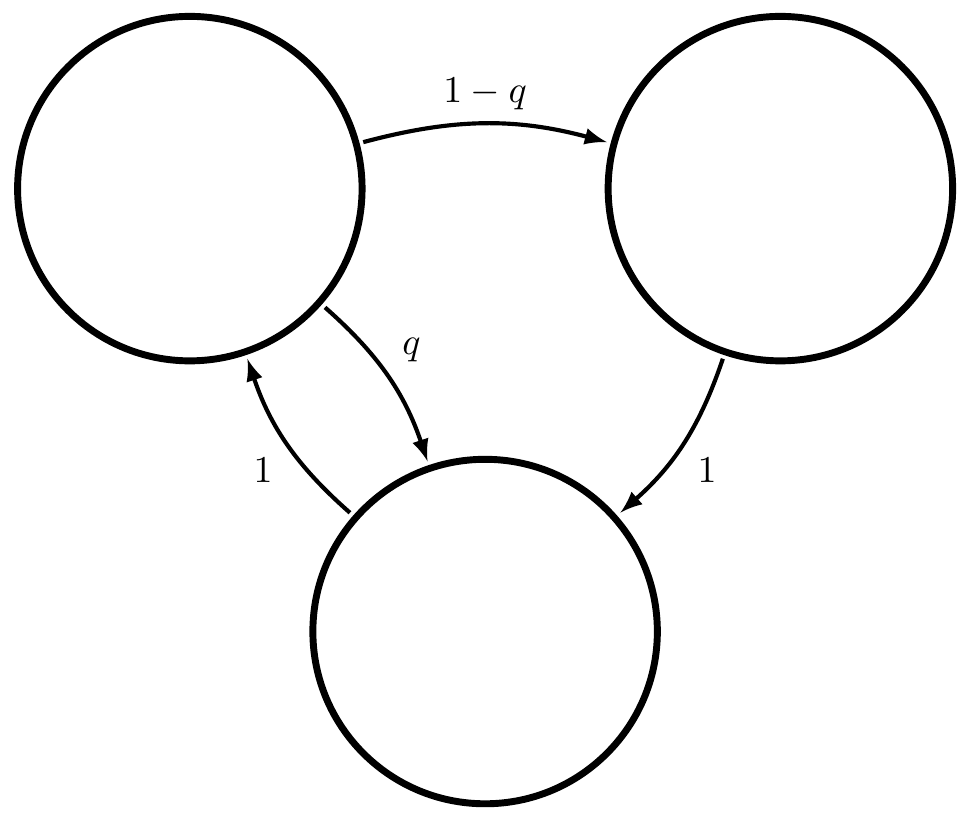}
  \put(45,38){\includegraphics[width=0.105\textwidth]{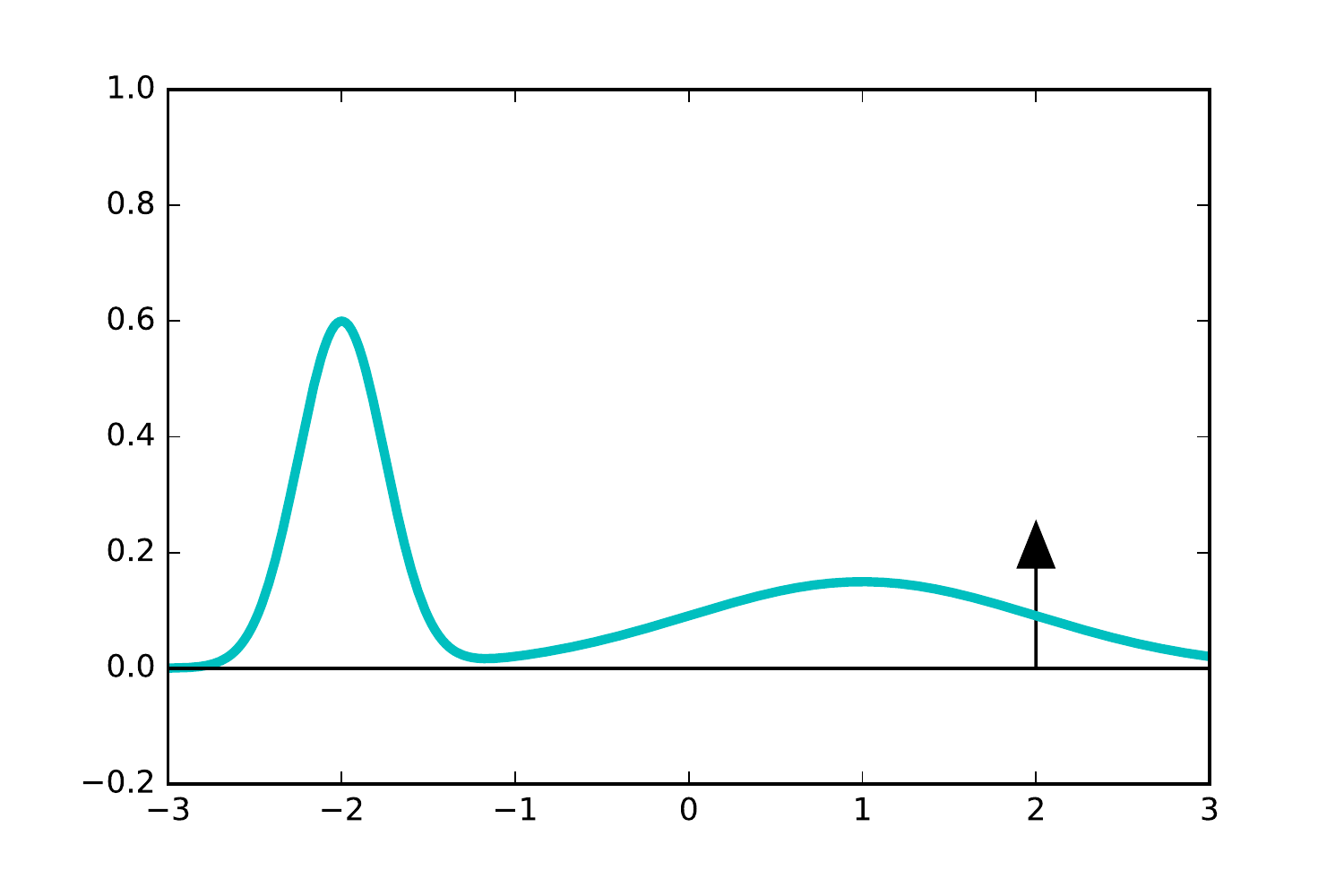}}
  \put(3.5,38){\includegraphics[width=0.105\textwidth]{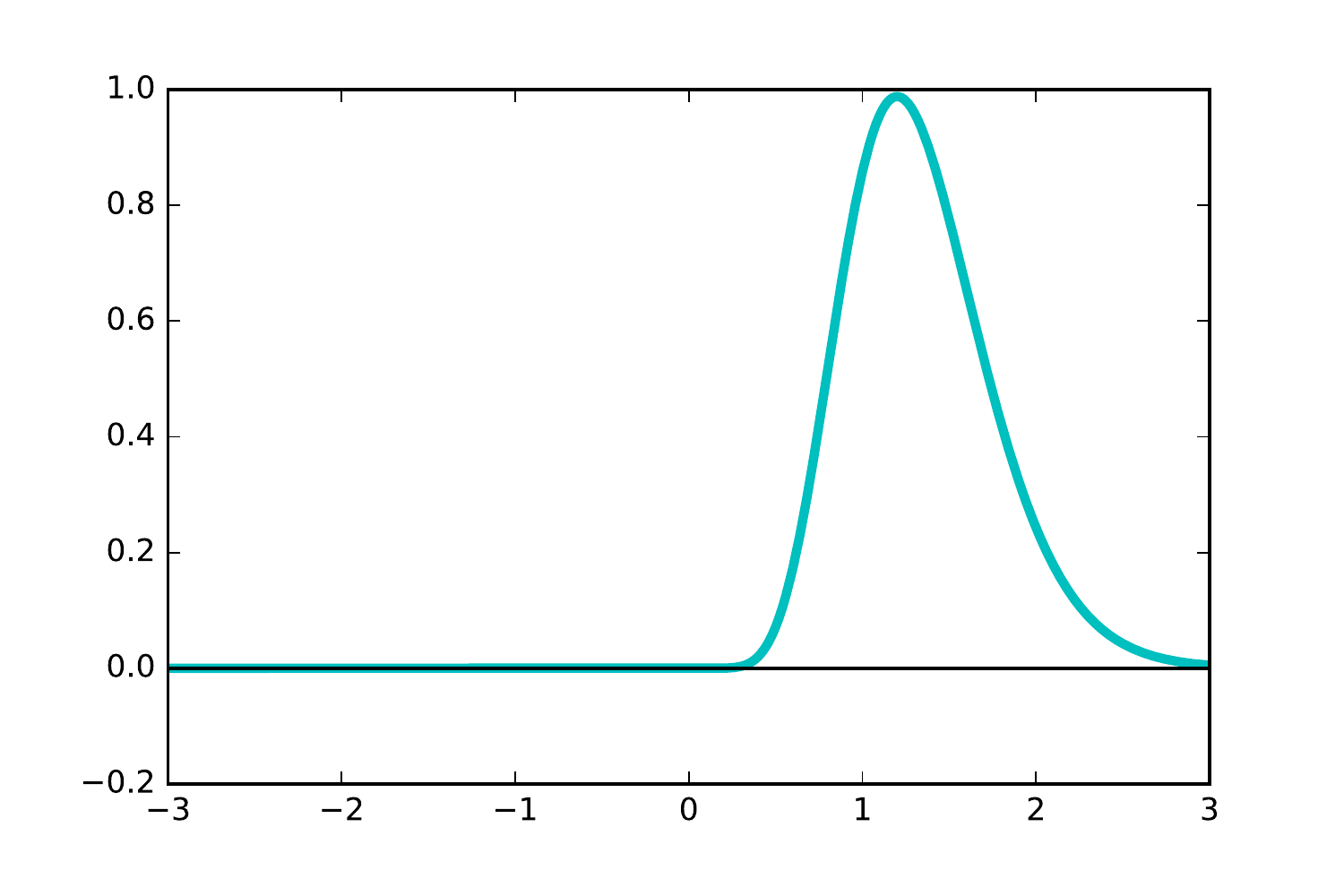}}
  \put(24,7.5){\includegraphics[width=0.105\textwidth]{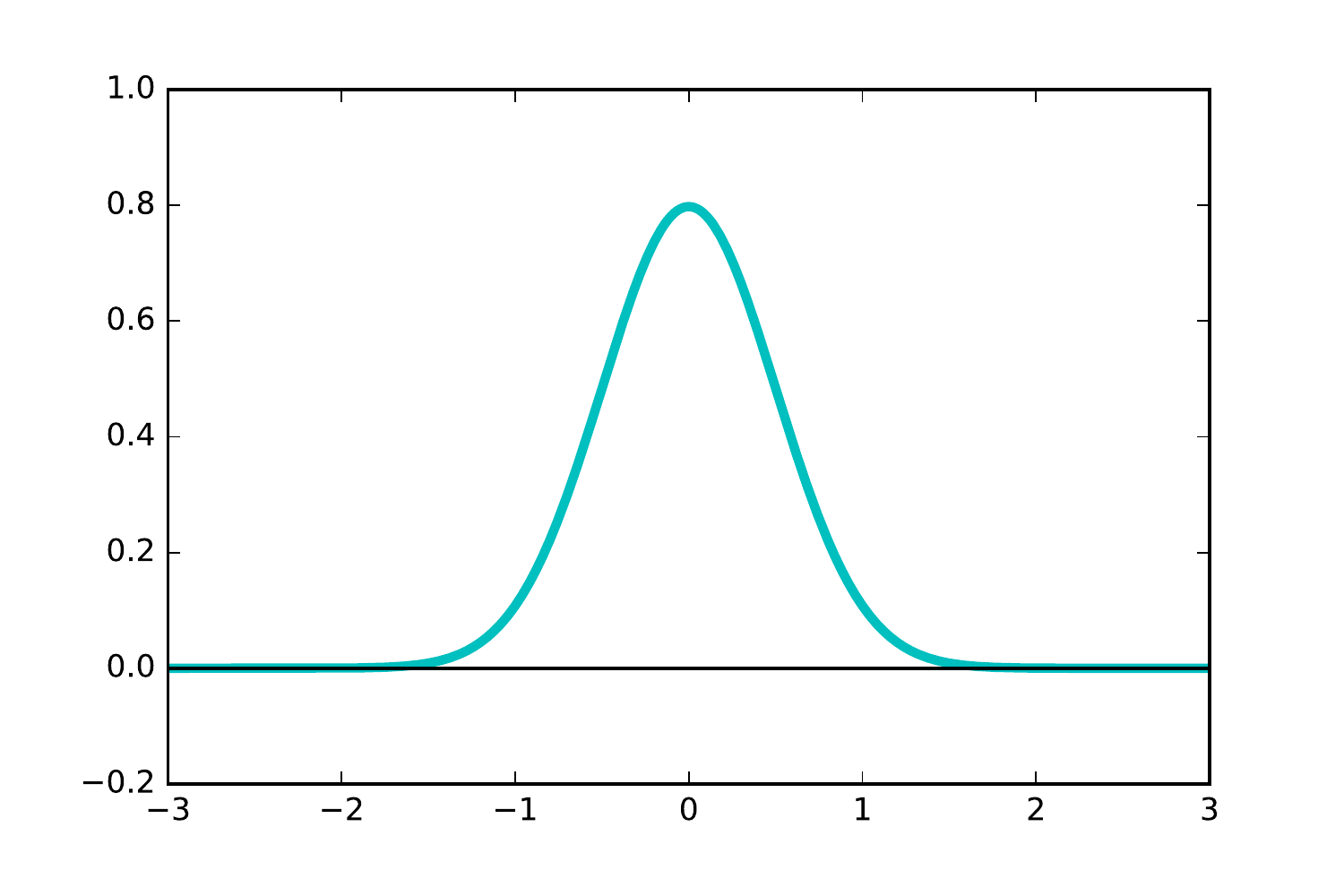}}
\end{overpic}
\end{center}
\caption{Simple $3$-state state-emitting HMM that generates a stochastic process
	according to the state-to-state transition dynamic $T$ and the probability
	density functions (pdfs) $\{ \pdf(x | s) \}_{s \in \SSet}$ associated with
	each state. Theorem 1 asserts that its power spectrum will be the same
	(with only constant offset) as the power spectrum generated from the
	alternative process where the pdfs in each state are solely concentrated at
	the Platonic average value $\langle x \rangle_{\pdf_s (x)}$ of the former
	pdf associated with the state.
	}
\label{fig:CC_HMM}
\end{figure}

Let the 4-tuple $\mathcal{M} = \bigl( \SSet , \Abet , \mathcal{P}, T  \bigr)$
be some discrete-time \emph{state-emitting hidden Markov model} (HMM) that
generates the stationary stochastic process $\dots X_{-2} X_{-1} X_0 X_1 X_2
\dots$ according to the following. $\SSet$ is the (finite) set of latent states
of the hidden Markov chain and $\Abet \subseteq \mathbb{C}$ is the observable
alphabet. $\St_t$ is the random variable for the hidden state at time $t$ that
takes on values $s \in \SSet$. $X_t$ is the random variable for the observation
at time $t$ that takes on values $x \in \Abet$. Given the latent state at time
$t$, the possible observations are distributed according to the conditional
probability density functions: $\mathcal{P} = \bigl\{ \pdf(X_t = x | \St_t = s)
\bigr\}_{s \in \SSet}$. For each $s \in \SSet$, $\pdf(X_t = x | \St_t = s)$ may
be abbreviated as $\pdf(x|s)$ since the probability density function in each
state is assumed not to change over $t$. Finally, the latent-state-to-state
stochastic transition matrix $T$ has elements $T_{i,j} = \Pr(\St_{t+1} = s_j |
\St_t = s_i)$, which give the probability of transitioning from latent state
$s_i$ to $s_j$ given that the system is in state $s_i$, where $s_i, s_j \in
\SSet$. It is important for the subsequent derivation that we use $\Pr(\cdot)$
to denote a probability in contrast to $\pdf(\cdot)$ which denotes a
probability \emph{density}. The Bayes network diagram of
Fig.~\ref{fig:HMM_BayesNet} depicts the structure of conditional independence
among the random variables.

\subsubsection{Continuous-value, discrete-state and -time processes}

Figure~\ref{fig:CC_HMM} gives a particular HMM with continuous observable
alphabet $\Abet = \mathbb{R}$ distributed according to the probability
density function shown within each latent state. Processes generated as the
observation of a function of a Markov chain can be of either finite or infinite
Markov order. (They are, in fact, \emph{typically} infinite Markov order in the
space of processes~\cite{Jame10a}.)

\begin{widetext}
Directly calculating, one finds that the autocorrelation function, for $\tau >
0$, for any such HMM is:
\begin{align*}
\gamma(\tau) 
&= \Braket{ \, \overline{X_n} X_{n + \tau}}_n \nonumber \\
&= \int_{x \in \Abet} \int_{x' \in \Abet} \overline{x} x' \pdf(X_0 = x, X_\tau = x') \, dx \, dx' \\
&= \sum_{s \in \SSet} \sum_{s' \in \SSet} \int_{x \in \Abet} \int_{x' \in
\Abet} \overline{x} x' \pdf(X_0 = x, X_\tau = x', \St_0 = s, \St_\tau = s') \, dx \, dx' \\
&= \sum_{s \in \SSet} \sum_{s' \in \SSet} \int_{x \in \Abet} \int_{x' \in \Abet} \overline{x} x' 
\Pr(\St_0 = s, \St_\tau = s') \, \pdf(X_0 = x | \St_0 = s ) \, \pdf( X_\tau = x' | \St_\tau = s') \, dx \, dx' \\
&= \sum_{s \in \SSet} \sum_{s' \in \SSet}  
\braket{\stationary | \delta_s} \bra{\delta_s} T^{\tau} \ket{\delta_{s'}} \braket{\delta_{s'} | \one} \, 
\Bigl( \int_{x \in \Abet} \overline{x} \, \pdf( x | s ) \, dx \Bigr)
\, \Bigl( \int_{x' \in \Abet} x' \, \pdf( x' | s' ) \, dx' \Bigr) \\
&= 
\bra{\stationary} \Bigl( \sum_{s \in \SSet} \braket{\overline{x}}_{\pdf( x | s )} \ket{ \delta_s} \bra{\delta_s} \Bigr) T^{\tau} 
\Bigl( \sum_{s' \in \SSet} \braket{x}_{\pdf( x | s' )} \ket{ \delta_{s'}} \bra{\delta_{s'}} \Bigr) \ket{\one} 
~,
\end{align*}
where:
\begin{align*}
\pdf (X_0 = x, X_\tau = x', \St_0 = s, \St_\tau = s')
  = \Pr(\St_0 = s, \St_\tau = s')
  \pdf(X_0 = x, X_\tau = x' | \St_0 = s, \St_\tau = s')
\end{align*}
holds by definition of conditional probability. The decomposition of:
\begin{align*}
\pdf(X_0 = x, X_\tau = x' | \St_0 = s, \St_\tau = s')
  = \pdf(X_0 = x | \St_0 = s ) \pdf( X_\tau = x' | \St_\tau = s') 
\end{align*}
for $\tau \neq 0$ follows from the conditional independence in the
relevant Bayesian network shown in Fig.~\ref{fig:HMM_BayesNet}.
Moreover, the equality:
\begin{align*}
\Pr(\St_0 = s, \St_\tau = s')
  = \braket{\stationary | \delta_s}
  \bra{\delta_s} T^{\tau} \ket{\delta_{s'}} \braket{\delta_{s'} | \one}
\end{align*}
can be derived by marginalizing over all possible intervening state sequences.
Note that $\ket{\delta_s}$ is the column vector of all $0$s except for a $1$ at
the index corresponding to state $s$ and $\bra{\delta_s}$ is simply its
transpose. Recall that $\bra{\stationary} = \bra{1_T}$ is the stationary
distribution induced by $T$ over latent states and $\ket{\one} =
\ket{1_T}$ is a column vector of all ones. 
Note also that $\braket{\stationary | \delta_s} = \Pr(s)$ and
$\braket{\delta_{s'} | \one} = 1$.
\end{widetext}

Since the autocorrelation function is symmetric in $\tau$ and:
\begin{align*}
\gamma(0) & = \bigl\langle \left| x \right|^2 \bigr\rangle_{\pdf(x)} \\
          & = \bra{\stationary}
		  \sum_{s \in \SSet} \bigl\langle \left| x \right|^2
		  \bigr\rangle_{\pdf(x|s)} \ket{\delta_s}
  ~,
\end{align*}
we find the full autocorrelation function is given by:
\begin{align*}
\gamma(\tau) & = 
\begin{cases}
\bigl\langle \left| x \right|^2 \bigr\rangle & \text{if } \tau = 0 \\
\bra{\stationary} \overline{\Omega} \, T^{|\tau| - 1} \Omega \ket{\one} & \text{if } |\tau| \geq 1
\end{cases}
~,
\end{align*}
where 
$\Omega$ is the $|\SSet|$-by-$|\SSet|$ matrix defined by:
\begin{align}
\Omega 
= \sum_{s \in \SSet} \braket{x}_{\pdf( x | s )} \ket{ \delta_s} \bra{\delta_s} T ~.
\end{align}

\begin{figure*}
\begin{subfigure}[t]{0.25\textwidth}
\centering
\includegraphics[width=0.8\textwidth]{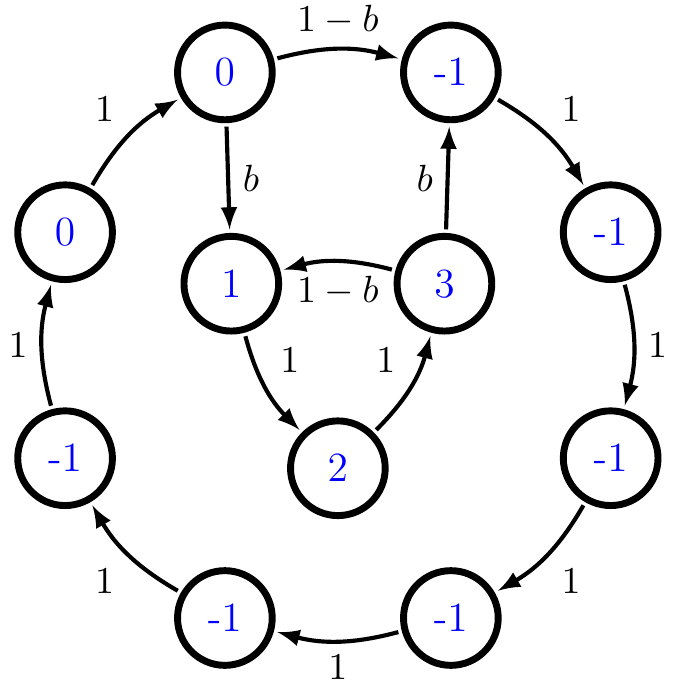}
\caption{A $b$-parametrized HMM with mean values of each state's pdf
	$\braket{x}_{\pdf(x|s)}$ indicated as the number inside each state.
	}
\label{fig:HMM4PSD}
\end{subfigure}%
    ~ 
\begin{subfigure}[t]{0.25\textwidth}
\centering
\includegraphics[width=0.95\textwidth]{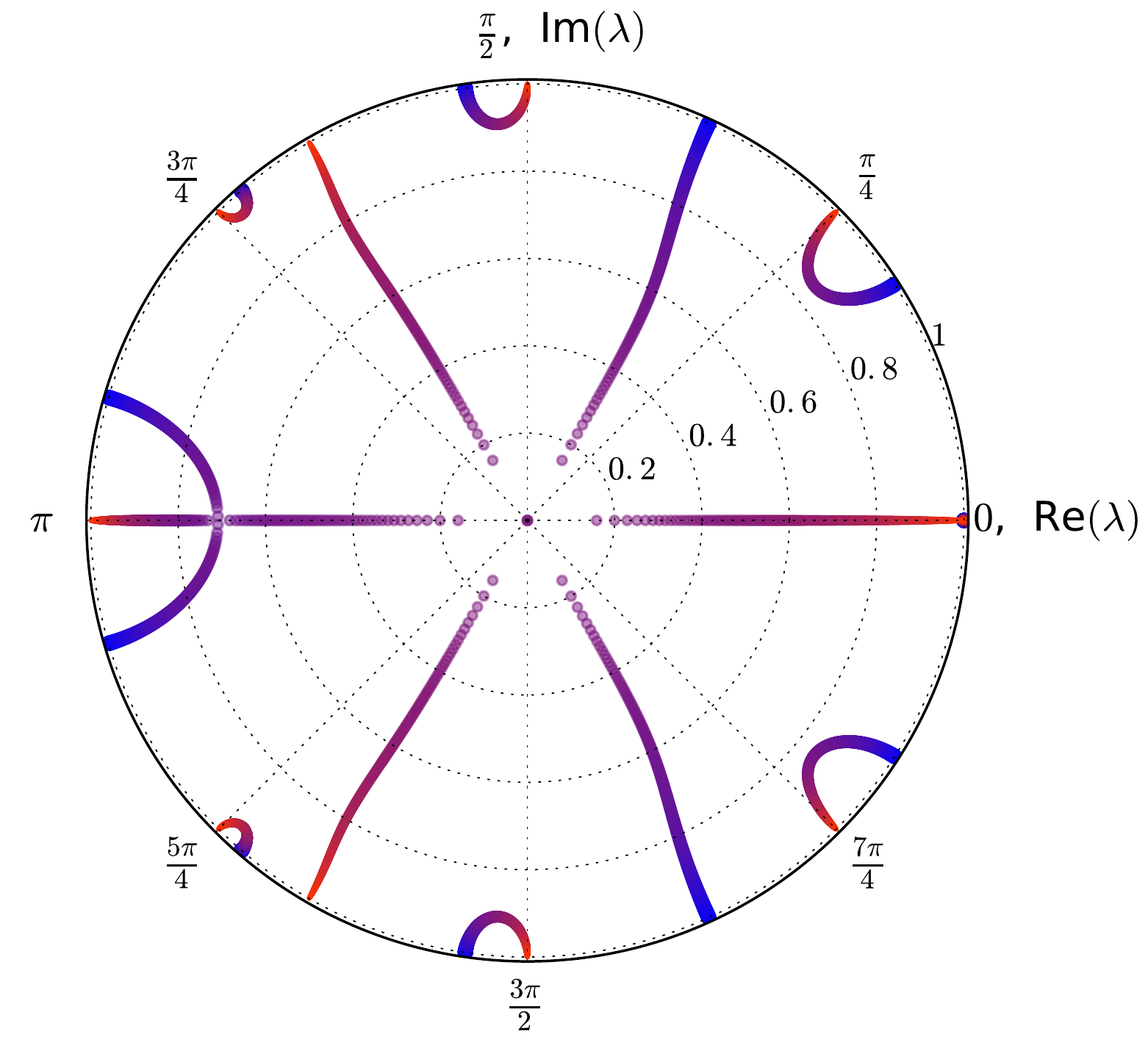}
\caption{Eigenvalue evolution for all $\lambda \in \Lambda_T$ sweeping 
	transition parameter $b$ from $1$ to $0$.
	}
\label{fig:HMMeigEvol}	
\end{subfigure}%
    ~ 
\begin{subfigure}[t]{0.25\textwidth}
\centering
\includegraphics[width=0.9\textwidth]{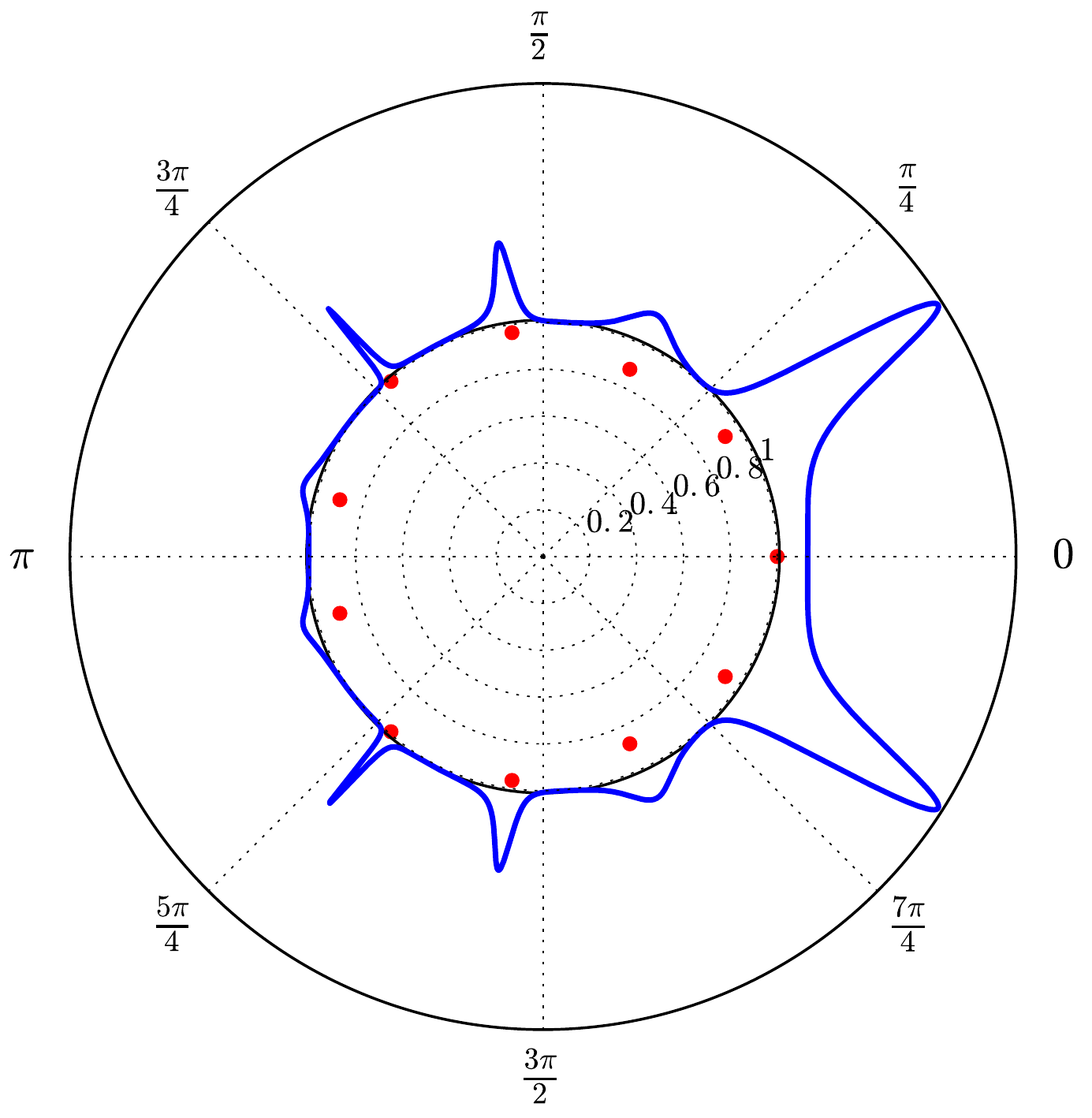}
\caption{Power spectrum and eigenvalues at $b = 3/4$.}
\label{fig:PSDfromSpectra1}
\end{subfigure}%
    ~ 
\begin{subfigure}[t]{0.25\textwidth}
\centering
\includegraphics[width=0.9\textwidth]{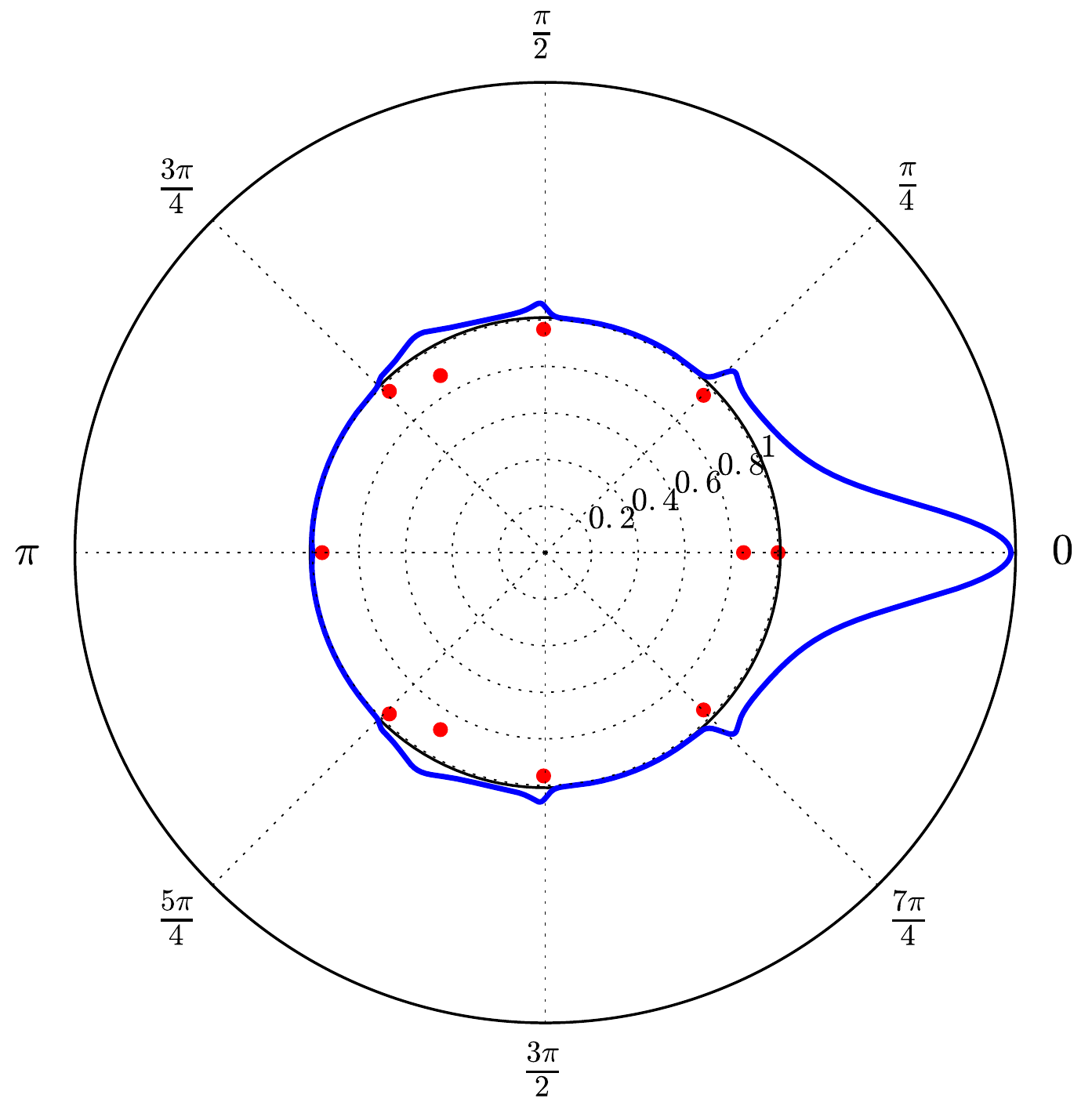}
\caption{Power spectrum and eigenvalues at $b = 1/4$.}
\label{fig:PSDfromSpectra2}
\end{subfigure}%
\caption{Parametrized HMM generator of a stochastic process, its eigenvalue
	evolution, and two coronal spectrograms showing power spectra emanating
	from eigen-spectra.
	}
\label{fig:CC_DP}
\end{figure*}

The power spectrum is then calculated via Eq.~\eqref{eq:PSDfromACF} using the
meromorphic calculus. In particular, the power spectrum decomposes naturally
into a discrete part and a continuous part. Full details will be given
elsewhere, but the derivation is similar to that given in Ref.~\cite{Riec14b}
for the special case of diffraction patterns from HMMs. We note that it is
important to treat individual eigenspaces separately, as our generalized
calculus naturally accommodates. The end result, for the continuous part of the
power spectrum, is:
\begin{align}
P_\text{c}(\omega) 
= \bigl\langle \left| x \right|^2 \bigr\rangle
 + 2 \, \text{Re} \bra{\stationary} \overline{\Omega} \, \bigl( e^{i \omega} I - T \bigr)^{-1} \Omega \ket{\one} ~.
\label{eq:PcwFromResolvent}
\end{align}
All of the $\omega$-dependence is in the resolvent. Using the spectral
expansion of the resolvent given by
Eq.~\eqref{eq:PartialFractionsExpansion_of_Resolvent_2} allows us to better
understand the qualitative possibilities for the shape of the power spectrum:
\begin{align}
P_\text{c}(\omega) = \bigl\langle \left| x \right|^2 \bigr\rangle
  + \sum_{\lambda \in \Lambda_T} \sum_{m = 0}^{\nu_\lambda - 1}
  2 \, \text{Re}
  \frac{\bra{\stationary} \overline{\Omega} \, T_{\lambda, m} \Omega \ket{\one}}{(e^{i \omega} - \lambda)^{m+1}}  ~.
\label{eq:PcwFromDecomposedResolvent}
\end{align}
Note that $\bra{\stationary} \overline{\Omega} \, T_{\lambda, m} \Omega
\ket{\one}$ is a complex-valued scalar and all of the frequency dependence now
handily resides in the denominator.

The discrete portion (delta functions) of the power spectrum is:
\begin{align}
P_\text{d}(\omega) &= \sum_{k = -\infty}^{\infty} 
  \sum_{\lambda \in \Lambda_T \atop |\lambda| = 1}
  2 \pi \, \delta( \omega - \omega_\lambda + 2 \pi k) 
  \nonumber \\
  & \qquad \qquad \qquad \times 
  \text{Re} \bigl( \lambda^{-1} \bra{\stationary} \overline{\Omega} \, T_\lambda \Omega \ket{\one} \bigr)
 ~,
\label{eq:PdwFromResolvent}
\end{align}
where $\omega_\lambda$ is related to $\lambda$ by $\lambda = e^{i
\omega_\lambda}$. An extension of the Perron--Frobenius theorem guarantees that
the eigenvalues of $T$ on the unit circle have index $\nu_\lambda = 1$.

When plotted as a function of the angular frequency $\omega$ around the unit
circle, the power spectrum suggestively appears to emanate from the
eigenvalues $\lambda \in \Lambda_T$ of the hidden linear dynamic. See
Fig.~\ref{fig:CC_DP} for the analysis of an example parametrized process and
the last two panels for this display mode for the power spectra.

Eigenvalues of $T$ \emph{on} the unit circle yield Dirac delta functions in the
power spectrum. Eigenvalues of $T$ \emph{within} the unit circle yield more
diffuse line profiles, increasingly diffuse as the magnitude of the eigenvalues
retreats toward the origin.  Moreover, the integrated magnitude of each
contribution is determined by projecting pairwise observation operators onto
the eigenspace emanating the contribution. Finally, we note that
nondiagonalizable eigen-modes yield qualitatively different line profiles.

Remarkably, the power spectrum generated by such process is the same as the
that generated by a potentially much simpler one---a process that is a function
of the same underlying Markov chain but instead emits the state-dependent
\emph{expectation value} of the observable within each state:

{\The \label{thm:PSDequivalence} 
Let $\mathcal{P} = \bigl\{ p_s(x) \bigr\}_{s \in \SSet}$ be any set of
probability distribution functions over the domain $\Abet \subseteq
\mathbb{C}$. Let $\mathcal{B} = \bigl\{ \braket{x}_{p_s(x)} \bigr\}_{s \in
\SSet}$ and let $\mathcal{Q} = \bigl\{ \delta(x - \braket{x}_{p_s(x)} )
\bigr\}_{s \in \SSet}$. Then, the power spectrum generated by any hidden Markov
model $\mathcal{M} = \bigl( \SSet , \Abet , \mathcal{P}, T  \bigr)$ differs at
most by a constant offset from the power spectrum generated by the hidden
Markov model $\mathcal{M}' = \bigl( \SSet , \mathcal{B} , \mathcal{Q}, T
\bigr)$ that has the same latent Markov chain but in any state $s \in \SSet$
emits, with probability one, the average value $\braket{x}_{p_s(x)}$ of the
state-conditioned probability density function $p_s(x) \in \mathcal{P}$ of
$\mathcal{M}$.
}

{\ProThe 
From Eqs. \eqref{eq:PcwFromResolvent} and \eqref{eq:PdwFromResolvent}, we see
that $P_\text{c}(\omega) + P_\text{d}(\omega) - \bigl\langle \left| x \right|^2
\bigr\rangle$ depends only on $T$ and $\bigl\{ \braket{x}_{\pdf(x|s)} \}_{s \in
\SSet}$. This shows that all HMMs that share the same $T$ and $\bigl\{
\braket{x}_{\pdf(x|s)} \}_{s \in \SSet}$ have the same power spectrum $P(\omega)
= P_\text{c}(\omega) + P_\text{d}(\omega)$ besides a constant offset determined
by differences in $\bigl\langle \left| x \right|^2 \bigr\rangle$.
}

One immediate consequence is that \emph{any hidden Markov chain with any
arbitrary set of zero-mean distributions attached to each state}, i.e.: 
\begin{align*}
\mathcal{P} \in \bigl\{ \{ \pdf(x | s) \}_{s \in \SSet} : \braket{x}_{\pdf(x|s)} = 0 \text{ for all } s \in \SSet \bigr\}
  ~,
\end{align*}
\emph{generates a flat power spectrum with the appearance of white noise}. On
the one hand, this strongly suggests to data analysts to look beyond power
spectra when attempting to extract a process' full architecture. On the other,
whenever a process's power spectrum \emph{is} structured, it is a direct
fingerprint of the resolvent of the hidden linear dynamic. In short, the power
spectrum is a filtered image of the resolvent along the unit circle.

The power spectrum of a particular stochastic process is shown in
Fig.~\ref{fig:CC_DP} and using \emph{coronal spectrograms}, introduced in Ref.
\cite{Riec14b}, it illustrates how the observed spectrum can be thought of as
emanating from the spectrum of the hidden linear dynamic, as all power spectra
must.  Figure~\ref{fig:HMM4PSD} shows the state-emitting HMM with
state-to-state transition probabilities parametrized by $b$; the mean values
$\braket{x}_{\pdf(x|s)}$ of each state's pdf $\pdf(x|s)$ are indicated as the
blue number inside each state. The process generated depends on the actual pdfs
and the transition parameter $b$ although, and this is our point, the power
spectrum is ignorant to the details of the pdfs.

The evolution of the eigenvalues $\Lambda_T$ of the transition dynamic among
latent states is shown from thick blue to thin red markers in
Fig.~\ref{fig:HMMeigEvol}, as we sweep the transition parameter $b$ from 1 to
0.  A subset of the eigenvalues pass continuously but very quickly through the
origin of the complex plane as $b$ passes through $1/2$. The continuity of this
is not immediately apparent numerically, but can be revealed with a finer
increment of $b$ near $b \approx 1/2$. Notice the persistent eigenvalue of
$\lambda_T = 1$, which is guaranteed by the Perron--Frobenius theorem.

In Fig.~\ref{fig:PSDfromSpectra1} and again, at another parameter setting, in
Fig.~\ref{fig:PSDfromSpectra2}, we show the continuous part of the \emph{power}
spectrum $P_\text{c}(\omega)$ (plotted around the unit circle in solid blue)
and the \emph{eigen}-spectrum (plotted as red dots on and within the unit
circle) of the state-to-state transition matrix for the $11$-state hidden
Markov chain (leftmost panel) that generates it. There is also a
$\delta$-function contribution to the power spectrum at $\omega = 0$
(corresponding to $\lambda_T = 1$). This is not shown. These coronal
spectrograms illustrate how the power spectrum emanates from the HMM's
eigen-spectrum, with sharper peaks when the eigenvalues are closer to the unit
circle. This observation is fully explained by
Eq.~\eqref{eq:PcwFromDecomposedResolvent}. The integrated magnitude of each
peak depends on $\bra{\stationary} \overline{\Omega} \ket{\lambda} \bra{\lambda} \Omega \ket{\one}$.

Interestingly, the apparent continuous spectrum component is the shadow of the
discrete spectrum of nonunitary dynamics. This suggests that resonances in
various physics domains concerned with a continuous spectrum can be modeled as
simple consequences of nonunitary dynamics. Indeed, hints of this appear in the
literature~\cite{Nare03,Soko06,Most09}.

\subsubsection{Continuous-time processes}

We close this exploration of conventional signal analysis methods using
the meromorphic calculus by commenting on continuous-time processes.
Analogous formulae can be derived with similar methods for 
continuous-time
hidden Markov jump processes and continuous-time deterministic (possibly chaotic)
dynamics in terms of the generator $G$ of time evolution.  
For example, 
the continuous part $P_\text{c}(\omega)$ of the power spectrum from a continuous-time deterministic dynamic has the form:
\begin{align*}
P_\text{c}(\omega) = 
 2 \, \text{Re} \bra{\stationary}
 \overline{\Omega} \, \bigl( i \omega I - G \bigr)^{-1} \Omega \ket{\one}
 ~.
\end{align*}
Appealing to the resolvent's spectral expansion again allows us to better
understand the possible shapes of their power spectrum:
\begin{align}
P_\text{c}(\omega) = 
\sum_{\lambda \in \Lambda_G} \sum_{m = 0}^{\nu_\lambda - 1}
  2 \, \text{Re} \frac{
  \bra{\stationary} \overline{\Omega} \, G_{\lambda, m} \Omega \ket{\one}}{(i \omega - \lambda)^{m+1}}
\label{eq:ContTimePSD}
  ~.
\end{align}
Since all of the frequency-dependence has been isolated in the denominator and
$ \bra{\stationary} \overline{\Omega} \, G_{\lambda, m} \Omega \ket{\one}$ is a
frequency-independent complex-valued constant, peaks in $P_\text{c}(\omega)$
can only arise via contributions of the form Re$\frac{c}{(i \omega -
\lambda)^n}$ for $c \in \mathbb{C}$, $\omega \in \mathbb{R}$, $\lambda \in
\Lambda_G$, and $n \in \mathbb{Z}_+$. This provides a rich starting point for
application and further theoretical investigation.  For example,
Eq.~\eqref{eq:ContTimePSD} helps explain the shapes of power spectra of
nonlinear dynamical systems, as have appeared, e.g., in Ref.~\cite{Farm80}.
Furthermore, it suggests an approach to the inverse problem of inferring the
spectrum of the hidden linear dynamic via power spectra. In the next section,
however, we develop a more general proposal for inferring eigenvalues from a
time series. Further developments will appear elsewhere.

\subsection{Operators for chaotic dynamics}

Since trajectories in state-space can be generated independently of each other,
any nonlinear dynamic corresponds to a linear operation on an
infinite-dimensional vector-space of complex-valued distributions 
(in the sense of generalized functions) over the original state-space. 
For example, the
well known Lorenz ordinary differential equations~\cite{Lore63a} are nonlinear
over its three given state-space variables---$x$, $y$, and $z$. Nevertheless,
the dynamic is linear in the infinite-dimensional vector space
$D(\mathbb{R}^3)$ of distributions over $\mathbb{R}^3$. Although
$D(\mathbb{R}^3)$ is an unwieldy state-space, the dynamics there might be well
approximated by a finite truncation of its modes.

\subsubsection{Ruelle--Frobenius--Perron and Koopman operators}
\label{sec:Ruelle_and_Koopman_operators}

The preceding operator formalism applies, in principle at least. The question,
of course, is, Is it practical and does it lead to constructive consequences?
Let's see.  The right eigenvector is either $\ket{0_G}$ or $\ket{1_T}$ with $T
= e^{\tau G}$ as the Ruelle--Frobenius--Perron transition
operator~\cite{Ruel71,Mack92a}. Equivalently, it is also $\boldsymbol{\pi}$,
the stationary distribution, with support on attracting subsets of
$\mathbb{R}^3$ in the case of the Lorenz dynamic. The corresponding
left-eigenvector $\one$, either $\bra{0_G}$ or $\bra{1_T}$, is uniform over the
space. Other modes of the operator's action, according to the eigenvalues and
left and right eigenvectors and generalized eigenvectors, capture the decay of
arbitrary distributions on $\mathbb{R}^3$.

The meromorphic spectral methods developed above give a view of the Koopman
operator and Koopman modes of nominally nonlinear dynamical
systems~\cite{Budi12} that is complementary to the Ruelle--Frobenius--Perron
operator. The Koopman operator $K$ is the adjoint---in the sense of vector
spaces, not inner product spaces---of the Ruelle--Frobenius--Perron operator
$T$: effectively the transpose $K = T^\top$. Moreover, it has the same spectrum
with only right and left swapping of the eigenvectors and generalized
eigenvectors.

The Ruelle--Frobenius--Perron operator $T$ is usually associated with the
evolution of probability density, while the Koopman operator $K$ is usually
associated with the evolution of linear functionals of probability density. The
duality of perspectives is associative in nature: $\bra{f} \bigl( T^n
\ket{\rho_0} \bigr)$ corresponds to the Ruelle--Frobenius--Perron perspective
with $T$ acting on the density $\rho$ and $\bigl( \bra{f} T^n \bigr)
\ket{\rho_0}$ corresponds to the Koopman operator $T^\top = K$ acting on the
observation function $f$. Allowing an observation vector $\vec{f} = [f_1, f_2 ,
\dots f_m]$ of linear functionals, and inspecting the most general form of
$K^n$ given by Eq.~\eqref{eq: T^n generally} together with the generalized
eigenvector decomposition of the projection operators of
Eq.~\eqref{eq:ProjectorsViaGenEigenvectors}, yields the most general form of
the dynamics in terms of Koopman modes. Each Koopman mode is a length-$m$
vector-valued functional of a Ruelle--Frobenius--Perron right eigenvector or
generalized eigenvector.

Both approaches suffer when their operators are defective. Given the
meromorphic calculus' ability to work around a wide class of such defects,
adapting it the Ruelle--Frobenius--Perron and Koopman operators suggests that
it may lift their decades-long restriction to only analyzing highly idealized
(e.g., hyperbolic) chaotic systems.



\subsubsection{Eigenvalues from a time series}

Let's explore an additional benefit of this view of the
Ruelle--Frobenius--Perron and Koopman operators, by proposing a novel method to
extract the eigenvalues of a nominally nonlinear dynamic. Let $O_N(f,z)$ be
($z^{-1}$ times) the $z$-transform \cite[pp. 257--262]{Brac99a} of a length-$N$
sequence of $\tau$-spaced type-$f$ observations of a dynamical system:
\begin{align*}
O_N(f,z) &\equiv z^{-1} \sum_{n=0}^N z^{-n} \braket{f |T^n| \rho_0} \\
  & \to_{N \to \infty}  \braket{f | (zI - T)^{-1} | \rho_0} \\
  & = \sum_{\lambda \in \Lambda_T} \sum_{m = 0}^{\nu_\lambda - 1}
  \frac{
  \bra{f}  T_{\lambda, m}  \ket{\rho_0}}{(r e^{i \omega} - \lambda)^{m+1}} 
  ~,
\end{align*}
as $N \to \infty$ for $ |z| = r > 1 $. Note that $\braket{f |T^n| \rho_0} $ is
simply the $f$-observation of the system at time $n \tau$, when the system
started in state $\rho_0$. We see that this $z$-transform of observations
automatically induces the resolvent of the hidden linear dynamic. If the
process is continuous-time, then $T = e^{\tau G}$ implies $\lambda_T = e^{ \tau
\lambda_G}$, so that the eigenvalues should shift along the unit circle if
$\tau$ changes; but the eigenvalues should be invariant to $\tau$ in the
appropriate $\tau$-dependent conformal mapping of the inside of the unit circle
of the complex plane to the left half complex plane. Specifically, for any
experimentally accessible choice of inter-measurement temporal spacing $\tau$,
the fundamental set of continuous time eigenvalues $\Lambda_G$ can be obtained
from $\lambda_G = \tfrac{1}{\tau} \ln \lambda_T$, where each $\lambda_T \in
\Lambda_T$ is extrapolated from $c / (re^{i \omega} - \lambda_T)^n$ curves fit
to $O_N(f,re^{i \omega})$ for $c \in \mathbb{C}$, large $N$, and fixed $r$.

The square magnitude of $O_N(f,z)$ is related to the power spectrum generated
by $f$-type observations of the system. Indeed, the power spectrum generated by
any type of observation of a nominally nonlinear system is a direct fingerprint
of the eigenspectrum and resolvent of the hidden linear dynamic. This suggests
many opportunities for inferring eigenvalues and projection operators from
frequency-domain transformations of a time series.

\section{Conclusion}
\label{sec:Conclusion}

The original, abstract spectral theory of normal operators rose to central
importance when, in the early development of quantum mechanics, the eigenvalues
of Hermitian operators were detected experimentally in the optical spectra of
energetic transitions of excited electrons. We extended this powerful theory by
introducing the meromorphic functional calculus, providing the spectral theory
of \emph{nonnormal} operators. Our straightforward examples suggest that the
spectral properties of these general operators should also be experimentally
accessible in the behavior of complex---open, strongly interacting---systems.
We see a direct parallel with the success of the original spectral theory of
normal operators as it made accessible the phenomena of the quantum mechanics
of closed systems. This turns on nondiagonalizability and appreciating how
ubiquitous it is.

Nondiagonalizability has consequences for settings as simple as counting, as
shown in \S~\ref{sec:Poisson}. Moreover, there we found that
nondiagonalizability can be robust. The Drazin inverse, the negative-one power
in the meromorphic functional calculus, is quite common in the nonequilibrium
thermodynamics of open systems, as we showed in \S~\ref{sec:NoneqThermo}.
Finally, we showed that the spectral character of nonnormal and
nondiagonalizable operators manifests itself physically, as illustrated by
Figs.~\ref{fig:PSDfromSpectra1} and~\ref{fig:PSDfromSpectra2} of
\S~\ref{sec:PSD}.

From the perspective of functional calculus, nonunitary time evolution, open
systems, and non-Hermitian generators are closely related concepts since they
all rely on the manipulation of nonnormal operators. Moreover, each domain is
gaining traction. Nonnormal operators have recently drawn attention, from the
nonequilibrium thermodynamics of nanoscale systems~\cite{Gard16} to large-scale
cosmological evolution~\cite{Berk04}. In another arena entirely, complex
directed networks \cite{Newm10} correspond to nonnormal and
not-necessarily-diagonalizable weighted digraphs. There are even hints that
nondiagonalizable network structures can be optimal for implementing certain
dynamical functionality~\cite{Nishi06}. The opportunity here should be
contrasted with the well established field of spectral graph
theory~\cite{Chun97} that typically considers consequences of the spectral
theorem for normal operators applied to the symmetric (and thus normal)
adjacency matrices and Laplacian matrices. It seems that the meromorphic
calculus and its generalized spectral theory will enable a \emph{spectral
weighted digraph theory} beyond the purview of current spectral graph theory.

Even if the underlying dynamic is diagonalizable, particular questions or
particular choices of observable often \emph{induce} a nondiagonalizable hidden
linear dynamic. The examples already showed this arising from the simple
imposition of counting or assuming a Poissonian dynamic. In more sophisticated
examples, we recently found nondiagonalizable dynamic structures in quantum
memory reduction~\cite{Riec16a} and classical complexity measures
\cite{Crut13a}.

Our goal has been to develop tractable, exact analytical techniques for
nondiagonalizable systems. We did not discuss numerical implementation of
algorithms that naturally accompany its practical application. Nevertheless,
the theory does suggest new algorithms---for the Drazin inverse, projection
operators, power spectra, and more. Guided by the meromorphic calculus, such
algorithms can be made robust despite the common knowledge that numerics with
nondiagonalizable matrices is sensitive in certain ways.

The meromorphic calculus complements attempts to address nondiagonalizability,
e.g., via pseudospectra~\cite{Tref97, Tref05}. It also extends and simplifies
previously known results, especially as developed by Dunford~\cite{Dunf54a}.
Just as the spectral theorem for normal operators enabled much theoretical
progress in physics, we hope that our generalized and tractable analytic
framework yields rigorous understanding for much broader classes of complex
system. Importantly, the analytic framework should enable new \emph{theory} of
complex systems beyond the limited purview of numerical investigations.

While the infinite-dimensional theory is in principle readily adaptable from
the present framework, special care must be taken to guarantee a similar level
of tractability and generality. Nevertheless, even the finite-dimensional
theory enables a new level of tractability for analyzing
not-necessarily-diagonalizable systems, including nonnormal dynamics. Future
work will take full advantage of the operator theory, with more emphasis on
infinite-dimensional systems and continuous spectra.
Another direction forward is to develop creation
and annihilation operators within nondiagonalizable dynamics. In the study of
complex stochastic information processing, for example, this would allow
analytic study of infinite-memory processes generated by, say, stochastic
pushdown and counter automata \cite{Crut89e,Trav11b,Marz15a,Crut15a}. In a
physical context, such operators may aid in the study of open quantum field
theories. One might finally speculate that the Drazin inverse will help to tame
the divergences that arise there.

\section*{Acknowledgments}

JPC thanks the Santa Fe Institute for its hospitality. The authors thank John
Mahoney, Sarah Marzen, Gregory Wimsatt, and Alec Boyd for helpful discussions. We especially thank Gregory Wimsatt for his assistance with \S~\ref{sec:ProjOpsFromGenEigvects}.  This material is
based upon work supported by, or in part by, the U. S. Army Research Laboratory
and the U. S. Army Research Office under contracts W911NF-12-1-0234,
W911NF-13-1-0390, and W911NF-13-1-0340.

\bibliography{chaos}

\end{document}